\documentclass[useAMS,usenatbib]{mn2e}
\usepackage{graphicx}
\usepackage{threeparttable}

\title[Low-Mass Stars and  Brown Dwarfs in Orion]{An optical  spectroscopic  H-R diagram for low-mass stars and brown dwarfs in Orion}

\author[F. C. Riddick et al.]{F. C. Riddick$^{1,2}$, P. F. Roche$^{2}$, P. W. Lucas$^{3}$, \\
$^{1}$ Dept. of Astronomy \& Astrophysics, Penn State University, 525 Davey Lab, University Park, PA 16802, USA \\
$^{2}$ Astrophysics, University of Oxford, Dept of Physics, DWB, Keble Road Oxford OX1 3RH, UK \\
$^{3}$ Centre for Astrophysics,  University of Hertfordshire, College Lane, Hatfield, Herts, AL10 9AB, UK \\
}

\begin{document}

\date{Accepted XX Received 2007  XX; in original form 2007 April XX}


\maketitle


\begin{abstract}

The masses and temperatures of young low mass stars and brown dwarfs in star-forming regions are not yet well established because of uncertainties in the age of individual objects and the spectral type--temperature scale appropriate for objects with ages of only a few Myr.   
Using multi-object optical spectroscopy, 45 low-mass stars and brown dwarfs in the Trapezium Cluster in Orion have been classified and 44 of these confirmed as bona fide cluster members.  The spectral types obtained have been converted to effective temperatures using a temperature scale intermediate between those of dwarfs and giants, which is suitable for young pre-main sequence objects.  The objects have been placed on an H-R diagram overlaid with theoretical isochrones. The low-mass stars and the higher mass substellar objects are found to be clustered around the 1 Myr isochrone, while many of the lower mass substellar objects are located well above this isochrone.  An average age of 1 Myr is found for the majority of the objects.  Assuming coevality of the sources and an average age of 1 Myr, the masses of the objects have been estimated and range from 0.018--0.44 $M_{\sun}$.  The spectra also allow an investigation of the surface gravity of the objects by measurement of the sodium doublet equivalent width.  With one possible exception, all objects have low gravities, in line with young ages, and the Na indices for the Trapezium objects lie systematically below those of young stars and brown dwarfs in Chamaeleon, suggesting that the 820~nm Na index may provide a sensitive means of estimating ages in young clusters.   
\end{abstract}



\section{Introduction}
\subsection{Background}
\label{sec:introbg}
There has been enormous progress in brown dwarf (BD) research in the decade since the discovery of the first confirmed BDs and substellar objects have been discovered in their hundreds, both as single and multiple objects, as companions around stars in the local neighbourhood, in nearby open clusters, in star-forming regions (SFRs) and in major deep imaging surveys. 

Open clusters provide the great opportunity to study properties of a large group of objects with known age, metallicity and distance. The Trapezium Cluster, the inner 0.3 pc or 2 arcmin of the Orion Nebula Cluster (ONC) (Herbig \& Terndrup, 1986), is an ideal site for the study of star formation and that of low-mass stars and substellar objects in particular. Its extreme youth ( $\sim$ 1 Myr: see the discussion is section 6.1) enables the detection of the full mass spectrum down to very low mass objects, since they are substantially brighter than older objects of similar mass: according to model evolutionary tracks (e.g. Burrows et al. 1997) they are 3 orders of magnitude more luminous at an age of a few million years than they are at a few billion years. The cluster is also extremely richly populated and compact: the total membership of the ONC is 3500 stars and the total mass contained is 1800$M_{\sun}$ (O'Dell, 2001).  Its very high central density of $\sim$2x$10^{4}$ stars per cubic parsec (Hillenbrand \& Hartmann, 1998) allows photometry of several hundred sources in relatively small surveys (e.g. Hillenbrand 1997, 
Lucas \& Roche 2000, hereafter LR00) and makes it an ideal site for multi-object spectroscopy. The dense, dusty obscuring OMC-1 cloud ensures that background stars should appear only at very large reddenings (e.g. Hillenbrand \& Hartmann 1998),  and therefore most of the stars that appear associated with the ONC really are so. In addition, its proximity allows the detection of low-mass objects and also ensures that foreground contamination is low. The high galactic latitude (b= -19 deg) also minimises non-member contamination. The distance to Orion KL was measured at  437 $\pm$19\,pc  by Hirota et al.\ from the annual parallax of water maser emission, while most estimates of the distance to the Trapezium cluster lie in the range $400-480$\,pc.   In this paper we adopt a distance of 450 pc. 

Confirmation of substellarity, i.e. a mass below 0.075 $M_{\sun}$, is non-trivial since the spectral type of an object at the transition from stellar to substellar status (as well as that at the deuterium burning threshold) depends on the age of the object.  Since BDs have no long-lasting source of fuel,  deuterium-burning being relatively rapid, most of their evolution is dominated by a long, slow cooling process and they evolve through a series of spectral types as they cool, the types becoming later with age, so that a spectral type cannot be associated with a specific mass, unless the age of an object is known.  This may be so in a cluster if an average age can be assumed and assigned to all members.  

While photometric data yield estimates of effective temperature $(T_{\rm eff})$ and luminosity, spectroscopy allows an independent and hopefully much better estimate of $T_{\rm eff}$, and may provide indications of the surface gravity, crucial for providing evidence of cluster membership (see Section~\ref{membership}).  Spectroscopy may provide evidence of the presence of circumstellar disks which in turn indicate youth and hence membership of the cluster. If membership of a cluster can be confirmed for an object then the cluster age and distance can be assigned to it and the validity of young theoretical isochrones can be tested and the spectra can be used to test and improve model atmospheres.  

There are limitations however: while optical spectroscopy provides the best studied sets of spectral lines for classification, the visual extinction in a SFR severely limits the number of young BDs that can be observed in the far red optical spectral region.  Additionally, comparisons between observations and models for young objects in SFRs are more uncertain because of the large amount of extinction due to the surrounding dust which modifies both the intrinsic magnitude and the colours and affects the spectra.  

The aim of this project is to classify a large sample of low-mass stars and BDs in the Trapezium Cluster in Orion by an analysis of their red optical spectra using the methods of Riddick, Roche \& Lucas (2007; hereafter Paper I). The spectral types are converted to $T_{\rm eff}$ by the use of a suitable temperature scale, which allows an estimate of the average age of the cluster, by a comparison with theoretical evolutionary tracks on a Hertzsprung-Russell diagram (HRD). Under the assumption of a coeval population, masses will be estimated. 


\subsection{Outline}
In Section 2 we describe the data selected for observation, the experimental setup, the slit mask design, the observations and data reduction, including dereddening and nebular emission subtraction.
In Section~3 the classification is carried out according to the procedures described in Paper I.
In Section~4 the membership probability of the objects is determined, also using the procedures described in Paper I.
In Section~5 the HRD is constructed, including the conversion of spectral type to temperature using an appropriate temperature scale.  
In Section~6, the HRD is analysed and compared with other studies. 
In Section~7, we will summarise the results.

\section{Observations in the Trapezium Cluster}
\label{obs}

Observations were conducted on the 3.9m Anglo-Australian Telescope (AAT)  using the Taurus focal reducer  in multi-object spectroscopic mode and on the 8.0m Gemini-North telescope where the Gemini Multi-Object Spectrograph (GMOS) was used.  Time was awarded for four separate observing runs for the spectroscopy, two on each telescope. The observations are summarised in Table~1.

\begin{center}
\begin{table*}
\caption{Observation Log}
\label{obs-table}
\begin{tabular*}{380pt}{llccccc}

\hline
Telescope & Dates (2002) & Instrument   & Exposure & Slit Width  & Telluric Standards\\

& &  & Time (mins) & (arcsec) &   \\
\hline

AAT & Jan20 & Taurus & 60,20,15 & 1.5  & w485a\\

AAT & Dec 27-29 & Taurus   & 340,220 & 1.3  & HD84937 \\

Gemini-N & Mar (8 nights) & GMOS   & 240,200 & 0.8 &  Feige 34 \\

Gemini-N & 3,8,9,10 Nov; 2 Dec & GMOS  & 240,120 & 0.8 & GD71, HZ2 \\

\hline
\end{tabular*}
\end{table*}
\end{center}

\subsection{Selection of Targets}
The target objects for spectroscopy were chosen from the IJH survey of LR00 and from the JHK survey of Lucas et al. (2005). Targets were selected in areas away from the cluster's very bright nebulous core, to enable on average fainter objects to be observed in regions of lower background emission.

Objects were chosen  with magnitudes in the range  H=11-18, i.e. from low-mass stars down to planetary-mass object (PMO) candidates, but with most targets close to the H-burning limit at H$\sim$ 13.5 mag for lightly reddened objects. For all runs, sources were selected preferentially if they had low extinction, with $A_{V}$ $<$ 5, though some sources with higher extinction were included to fill in space in the masks which could not be filled by low-extinction objects. Several of the objects were observed through more than one mask and/or on more than one run to fill up available space, allowing a limited study of variability in a few  of the objects -- see Section~\ref{duplicates}.  The faintest objects included in the masks yielded  spectra with insufficient signal to noise ratio (S/N) to allow classification, so for the second runs at each telescope all but the brightest PMO  candidates were excluded 

\subsection{Slit Masks}
To maximise the number of objects observed, slit heights in the multi-object slit masks were 5-7 arcsec -- the minimum sufficient to allow the spectra at  the two telescope nod positions to lie within the height of the slits. Short nods (3-4 arcsec) were used to minimise the difference in nebular background after subtraction of neighbouring image pairs.  The AAT masks each contained 30-32 slitlets and the GMOS masks 23-26.

\subsection{AAT Observations using Taurus}
\label{taurus}
The AAT observations were carried out by FCR and PWL using Taurus++ at the f/8 Cassegrain focus of the telescope.  The  field in the vertical direction was 8.4 arcmin with a scale of 0.37 arcsec/pixel. A red volume phase holographic grating dispersed the light from multi-slit masks inserted at the telescope focal plane. An OG515 filter was used to avoid second order contamination of the spectra.  The detectors used were MITLL2A and MITLL3A for the 1st and 2nd runs respectively. Spectral coverage was 6000--10,000\,\AA~  for objects located within the central 2 arcmin of the field in the dispersion direction, with reduced spectral coverage for objects located towards the edges. The resolving power employed  was R$\sim$ 1000, sampled at 2.8\,\AA/pixel.

Taurus was used in `nod and shuffle' mode (Bland-Hawthorn \& Jones 1998; Glazebrook \& Bland-Hawthorn 2001), the major advantage of which is extremely accurate subtraction of night sky emission in spectroscopy, since OH airglow changes on short timescales of a few minutes. It allows relatively rapid nodding of the telescope, providing good subtraction of the airglow lines with no penalty of multiple CCD readouts, resulting in a high observing efficiency. In this case, cycles of 1 minute were used.  

Nod and shuffle did work very well for the subtraction of the spatially uniform though temporally variable terrestrial sky background but it could not completely subtract the highly spatially structured nebular background, which varies on spatial scales smaller than the nod length used.  The nebular spectrum, therefore, still often appeared as a series of lines in emission or absorption, depending on the local gradient in the extracted spectra.  This is especially true for $H\alpha$ emission, which was present and strong in all the spectra. 

Due to their intrinsic faintness, acquisition of the targets was not easy. Stability problems in the mask wheel (not rotating to exactly the expected positions) and a lack of suitable astrometric software made it very difficult to acquire the masks and for the first run only one mask was able to be observed for the full hour intended and only 3 masks were used in total, due in addition to time lost to poor weather. It was decided to observe only one mask on each night of the second run to minimise the acquisition overheads. 

\subsection{Gemini-North Observations using GMOS}
GMOS on Gemini-North was used for multi-object spectroscopy, providing a 5.5 arcmin FOV.  The R400 grating was used, giving spectral coverage from 6000--10,000\,\AA\, and a spectral resolving power of R$\sim$ 2000 and a sampling of $\sim$0.7\,\AA/pixel. The observations were carried out by observatory staff. The detector consisted of 3 CCDs each of size 2048x4608 pixels, i.e. a total of 6144 x 4608 pixels in the spectral and spatial directions respectively,  with a scale of 0.073 arcsec per pixel; see table 1.  Nod and shuffle was not available on GMOS and sky correction relied on subtraction from the nodded position. 

\subsection{Data Reduction}
The data were reduced using standard procedures within IRAF. Dispersed images were coadded before sky subtraction and extraction. The data were bias corrected, flat-fielded and corrected for telluric absorption. A telluric standard was observed on each run, to allow the spectra to be corrected for atmospheric absorption features.   Telluric bands should not present significant problems with spectral typing (e.g. Kirkpatrick et al. 1991), and the typing procedure used was designed to be relatively unaffected by the presence of telluric absorption.  Observations of standard stars were undertaken  at similar at similar airmasses to the average of those for the science observations. CuAr arc lamp spectra were obtained and used for wavelength calibration, typically accurate to 0.5 pixel. 

\subsubsection{Nebular Emission Line Extraction}
Nebular line subtraction is necessary since a significant source of noise arises from the spatial gradients in the nebular emission, which makes background subtraction problematic. We fit a polynomial to the residual background at each wavelength and subtracted this during the extraction, as done by Lucas et al. (2001, hereafter LR01). This was partially successful in removing nebulosity gradients (and residual OH emission), but it is not a perfect correction in most cases and nebular lines are still seen in many of the spectra even after sky subtraction. It is worst for the faintest sources which have peak brightness comparable to nebular surface brightness fluctuations on a scale of a few arcsec even in low background regions. 

\subsubsection{Dereddening of Spectra and Fluxes}
\label{dered}
In a young cluster such as Orion, it is necessary to correct for differential reddening towards individual stars since the extinction varies on small scales due to the highly structured nebulosity.  
The DEREDDEN task in IRAF was used to deredden the spectra to the theoretical NextGen 
(J-H) vs H colour-magnitude sequence of Baraffe et al. (1998; BCAH98 hereafter) at age 1 Myr. 
The reddening parameter was set at the typical interstellar value of $R_V=3.1$ since this
produces very similar extinction values to the Rieke \& Lebofsky (1985) extinction law,
which appears to be suitable for the Trapezium cluster (Lucas et al. 2005).

Uncertainties in A$_{V}$ of $\pm$ 1 mag are expected due to photometric errors and since many of the young sources may have IR colours affected substantially by emission from circumstellar disks and envelopes (Luhman \& Rieke, 1999).  Objects with negative derived extinction estimates are given a value of zero.

The procedure used for spectral typing was designed to ensure that the results are  relatively 
immune to the effects of reddening on the spectra and, as described in Paper I, errors in 
reddening corrections are likely to give errors in the spectral type of typically less than 0.1 
subtypes.

\section{Spectral Classification}
\label{class}
Young, low-mass objects are located on primarily vertical tracks on the HRD (e.g. BCAH98) therefore mass estimates are much more sensitive to errors in temperature than luminosity. Temperature is derived from the spectral type so it is crucially important to assign spectral types as accurately as possible. To this end, we devised in Paper I a classification system valid for the spectral range M3-M9 and independent of both reddening and nebular emission lines and we have used this system to classify these data.
Specifically, we employed the indices recommended in paper I, sampling the spectra near 7445, 8000, and 8440\,\AA~  using the VO 7445 index defined by Kirkpatrick et al (1991), the VO 2 index of  Lepine et al (2003), and the c81 index of Stauffer et al (1999).

\subsection{Spectral Types Assigned}
Following the spectral typing, 45 of the objects observed were able to be assigned a spectral type -- see Table~2. An additional 11 spectra were untypable due to low S/N or veiling but appeared to have M type features -- see Table~3. The observed and dereddened fluxes given in Table 2 use data 
from the UKIRT IJH photometry of LR00 and LR01 where good quality data are available, since these 
fluxes were obtained almost simultaneously. Fluxes from the Gemini data of Lucas et al. (2005) were 
used otherwise, except in the case of 148-831 (2MASS J05351475-0528318) where the only available 
data was the 2MASS fluxes. We note that the dereddened UKIRT H band magnitudes used in
the H-R diagrams agree with the Gemini data (within the error bars) in the great majority of cases.
The spectral types given by the indices are listed in the Appendix, but the final class assignations were done manually for this sample to ensure that spurious results were not introduced by nebular contamination. The spectra of all of the objects successfully extracted are shown in Figure A-1 in the Appendix and the final spectral classes assigned are listed in table 2.  Comparison with the types listed in the Appendix demonstrates that the final classes are generally consistent with those given by the indices within the estimated uncertainties, although in a few objects the indices give significantly different classes, demonstrating that using several indices covering a range of spectral features is desirable, in addition to an independent classification by comparison with other classified objects. 

\begin{center}
\begin{table*}
\caption{\textbf{Photometry \& Spectral Types Assigned}. Observed Cousins I band fluxes and 
dereddened I, J and H fluxes are given.}  
\label{types}
\begin{tabular*}{440pt}{llllllllll}

\hline

NAME & I$_{obs}$ & I$_{dr}$ & J$_{dr}$ & H$_{dr}$ & A(V) & Spectral Type & Type Error & 
ID(LR00) & ID(LRT05)\\
\hline

011-027 & 17.10 &	16.95	& 15.48 & 14.95 & 0.3  & 3.25	& 1.0 & 011-029 & \\
4584-117 & 15.77 &	13.62	& 12.08 & 11.37 & 3.7  & 3.5	& 0.5 & 4585-118 & \\
177-541 & 18.97 &	16.79	& 15.24 & 14.70 & 3.8  & 4.25	& 1.5 & 177-540 & 154\\ 
112-532 & 16.76 &	14.69	& 12.98 & 12.36 & 3.6  & 4.75	& 0.5 & 112-531 & 171 \\ 
082-403 & 15.95 &	13.59	& 12.23 & 11.53 & 4.1  & 4.75	& 0.5 & 083-403 & 265 \\
091-017 & 16.53 &	16.53	& 14.19 & 13.68 & 0.0  & 5.0	& 0.5 & 091-019 & \\
034-610 & 16.18 &	15.86   	& 13.81 & 13.24 & 0.6  & 5.25	& 0.5 & 035-609 & 117 \\
016-534 & 15.79 &	15.79	& 13.42 & 13.04 & 0.0  & 5.25	& 0.5 & 016-533 & 166 \\ 
130-458 & 17.09 &	14.78	& 13.01 & 12.38 & 4.0  & 5.25	& 0.5 & 130-458 & 220 \\
121-434 & 17.10 &	14.52	& 13.26 & 12.66 & 4.4  & 5.5 & 1.0 & 122-434 & \\
017-636 & 16.91 &	14.70   	& 13.13 & 12.51 & 3.8  & 5.5	& 1.0 & 017-635 & 79 \\
037-246 & 16.94 &	14.97	& 12.33 & 11.63 & 3.4  & 5.5	& 1.0 & 038-246 & 339 \\
222-745 & 	&			& 15.46 & 14.93 & 2.3  & 5.75	& 0.5 &  & 2\\
068-019 & 17.11 &	16.76	& 14.28 & 13.73 & 0.6  & 5.75	& 0.5 & 068-020 & \\
019-354 & 17.94 &	17.05	& 14.36 & 13.81 & 1.5  & 5.75	& 0.5 & 019-354 & 281 \\
103-157 & 15.98&       15.38	& 13.33 & 12.73 & 1.0  & 5.75	& 0.5 & 103-158 & 390 \\
017-710 & 16.45 &	16.32	& 13.76 & 13.18 & 0.2  & 6.0	& 0.5 & 018-709 & 36\\
069-209 & 16.32 &	16.32	& 14.09 & 13.67 & 0.0  & 6.0	& 0.5 & 069-210 & 380\\
156-547 & 19.59 &	18.60	& 16.34 & 15.82 & 1.7  & 6.0	& 1.0 & 156-546 & 149\\
4559-109 & 18.85 &	16.79	& 14.34 & 13.79 & 3.6  & 6.25	& 0.5 & 4560-109 &\\
102-102 & 16.95 &	16.95	& 14.32 & 13.86 & 0.0  & 6.25	& 0.5 & 102-102 & \\
095-058 & 16.51 &	16.51   	& 13.71 & 13.25 & 0.0  & 6.25	& 0.5 & 095-100 &\\
077-453 & 16.23 &	15.11	& 13.36	& 12.76 & 1.9  & 6.5	& 0.5 & 077-452 & 229\\
072-638 & 18.11 &	17.46   	& 15.20	& 14.66 & 1.1  & 6.5	& 0.5 & 073-637 & 70\\
053-503 & 17.46 &	16.25	& 13.64 & 13.06 & 2.1  & 6.5	& 1.0 & 053-503 & 208 \\
154-600 & 16.97 &	16.44   	& 13.48	& 12.89 & 0.9  & 6.5	& 0.5 & 154-559 & 132\\
014-413 & 18.40 &	17.47	& 14.71 & 14.16 & 1.6  & 6.5	& 1.0 & 014-413 & 254\\
4569-122 & 17.78 &	16.81	& 14.30 & 13.74 & 1.6  & 6.75	& 1.0 & 4570-123 &\\
035-333 & 16.84 &	16.84	& 13.81 & 13.24 & 0.0  & 6.75	& 0.5 & 035-333 & 290\\
096-1943 & 18.78 &	18.07	& 15.29 & 14.75 & 1.2  & 6.75 & 0.5 & 096-1944 & \\
055-230 & 17.68 &	17.15	& 13.69 & 13.11 & 0.9  & 6.75	& 1.5 & 055-231 & 358\\
084-305 & 18.01 &      15.70	& 13.12 & 12.50 & 4.0  & 7.0	& 0.5 & 085-305 & 320\\
030-524 & 20.59 &	20.11   	& 17.66 & 17.12 & 0.8  & 7.5     &1.5  & 031-524 & 178\\	
217-653 & 	& 			& 14.52 & 13.97 & 1.7  & 7.75	& 0.5 & & 55\\
042-012 & 18.12 &	17.62	& 14.32 & 13.76 & 0.9  & 7.75	& 1.0 & 043-014 & \\
092-606 & 20.74 &	19.90	& 16.68 & 16.17 & 1.4  & 8.0	& 2.0 & 092-605 & 122\\
186-631 & 19.54 &	18.84	& 15.60 & 15.07 & 1.2  & 8.0	& 1.0 & 186-631 & 85\\
130-053 & 20.24 &	19.98	& 15.65 & 15.12 & 0.4  & 8.5	& 2.0 & 131-054 & \\ 
047-550 & 20.89 &	20.30	& 16.96 & 16.45 & 1.0  & 8.5	& 2.0 & 047-549 & 143\\
077-127 & 18.76 &	18.24	& 14.69 & 14.14 & 0.9  & 8.5	& 1.0 & 078-128 & \\ 
082-253 & 19.88 &	18.01   	& 15.67	& 15.14 & 3.2  & 8.5 & 1.5 & 083-253 & 332 \\	
148-831 & 	&			& 15.53 & 15.00 & 4.8 & 8.5	& 2.0 &  & \\
183-729 & 	&		& 17.69 & 17.15 & 1.3 & 8.75	& 1.5 & & 18\\
165-634 & 20.21 &	18.79   & 15.70	& 15.17 & 2.4 & 8.75	& 1.0  & 166-634 & 82\\ 
031-536 & 20.23 &	20.23   & 16.72	& 16.23 & 0.0 & 8.75	& 1.0 & 032-536 & 159\\

\hline
\end{tabular*}
\end{table*}
\end{center}

\normalsize
\begin{center}
\begin{table*}
\caption{\textbf{Probable M-type Objects.} These showed basic M-type characteristics but were unclassifiable due to low contrast of spectral features or low S/N.} 
\label{notypes}
\begin{tabular*}{460pt}{lllllllllll}

\hline
NAME & I$_{obs}$ & I$_{dr}$ & J$_{dr}$ & H$_{dr}$ & A(V)$^*$ & Spectral Type & ID(LR00) & ID(LR05) & Notes \\
\hline
117-609 &     &	     & 10.08  & 9.51 & 12.2	& early-M & & 119 & veiled,CaII \\
050-143 & 20.62 & 20.25    & 17.57	&17.04 & 0.6	&mid-M & 050-144 & &  \\
196-700 & 21.09	&  19.23   & 16.95	&16.44 & 3.2	&mid-M & 196-659 & 47 &\\
046-245 & 19.89	&  19.18   & 16.92	&16.41 & 1.2	&late-M & 047-245 & 342 &\\	
139-425 & 17.65  & 16.55   & 14.49      & 13.94	& 1.9	&M & 140-425 & &\\
086-324 & 18.66  & 15.87   & 13.85	&13.27 & 4.8	&M & 086-324 & 296 &\\	
041-210 & 20.55	& 18.23    & 16.12	&15.60 & 4.0 &M & 042-210 & 379 &\\
160-607 &	 &    & 11.07	& 10.33 & 11.7	&M & & 120 & heavily veiled,\\
084-119 & 18.58  & 15.35  & 13.42	& 12.83 & 5.6	&M? & 085-121 & & CaII \\	
047-436 & 18.28	 & 14.67   & 13.85	& 13.28 & 6.2   &M & 047-436 & 242 & heavy veiled, CaII\\
044-219 &		& 15.65 & 15.12 &	  & 2.9  &  Late M\\

& & & & & & & \\

\hline
\end{tabular*}
\end{table*}
\end{center}


The names of the objects are derived from the coordinates, following the convention of O'Dell \& Wen (1994): the first 3 (or 4) digits are RA offset from 05 hours, 35 minutes and the last 3 (or 4) digits are declination offset from -05 degrees, 20 minutes.  For example, object 084-305 has RA = 05:35:08.4 and Dec = -05:23.05. 2MASS was used to calibrate the astrometry from the photometric surveys.  The coordinate based names used in LR00 and LR01 sometimes differ
slightly from those used here, since these were based on a less accurate astrometric calibration.
For clarity, Table 2 also includes the names used in those publications (ID LR00) and the running 
number used in Lucas et al.  (2005) is also given (ID LRT05).

\subsection{Veiling}
\label{veiling}

Robberto et al. (2004) find that since the ratio between the accretion
luminosity and the total stellar luminosity, corresponding to the
veiling factor of the stellar absorption lines, is generally small,
the positions of the stars in the H-R diagram remain almost unaffected
by the removal of any accretion contribution. Veiling of
narrow absorption lines can be high if hot gas in a
stellar wind produces an emission line that veils the stellar
absorption line, but this is likely to be less problematic  for broad
molecular bands.

Veiling in a spectrum from either dust scattering or circumstellar
emission can reduce the contrast of molecular absorption bands and
make a later type spectrum look like an earlier type.  The
effect is typically less than 1.5 subtypes (Hillenbrand 1997), but 
can occasionally be much larger.
A number of objects in this sample display  emission in the Ca II triplet lines indicative of
emission from  circumstellar material,  which could dilute molecular features.  
In most cases, any veiling appears to be weak, but  in two of the objects that could not 
be classified, the effect is prominent.  These  objects are described in section~\ref{disks} 
and the effects are discussed in section~\ref{slesnick}.

\subsection{Comparison of Spectral Types from Different Runs}
\label{duplicates}
To investigate variability of the objects and to check on the accuracy of spectral types obtained, we compared the spectral types obtained from different runs and found them to be generally in extremely good agreement -- see Table~4. Where the types do not agree, the types obtained from the GMOS runs were usually used for the final spectral type due to the higher S/N and resolution, but this was decided on a case by case basis.  The largest difference in spectral type for multiply observed objects is 0.75 subtypes and for types earlier than M6 the agreement is always within 0.5 subtypes.  Since the overall spectral typing error is 0.5 subtypes and 1--2 subtypes for types earlier and later than M6 respectively,  we conclude that we have not detected any significant spectral variations.   Note, however, that the spectrum of the unclassified object 047-436 does show significant changes; this is discussed in the Appendix. 

\begin{center}
\begin{table}
\caption{\textbf{Comparison of Spectral Types measured on more than 1 Observing Run}}
\label{comparison}
\begin{tabular}{llllll}
\hline

ID & AAT 1 & AAT 2 & GMOS 1 & GMOS 2 & Final Type \\
\hline

091-017 &              & M4.5   & M5         &              & M5 \\
034-610 & M5.25  &             &                & M5.25  & M5.25 \\
121-434 & M5.25  &	           &                & M5.75  & M5.5 \\
068-019 & M5.5    &	           & M5.75	      &              & M5.75 \\
019-354 & M5.5    &	           & M6	      &              & M5.75 \\
069-209 & M5.75  &	           & M6	      &              & M6 \\
095-058 & M6       &	           & M6.25    &              & M6.25 \\
053-503 & M6.5    &	           & M6	      &              & M6.5 \\
072-638 & M6       &	           &                & M6.75  & M6.5 \\
077-453 & M6.25  &	           &                & M6.5    & M6.5 \\
014-413 & M6.25  & M6.25 & M7          &             & M6.5 \\
035-333 & M6       & M6.25 & M6.75     &             & M6.75 \\
084-305 & M6.5    & M7.5   & M7.25     &             & M7 \\
042-012 & M8.5    &             & M7.75    &              & M7.75 \\
\hline
\end{tabular}
\end{table}
\end{center}


\subsection{Completeness of Spectroscopy}
\label{completeness}
\begin{figure}
\includegraphics[height=4in]{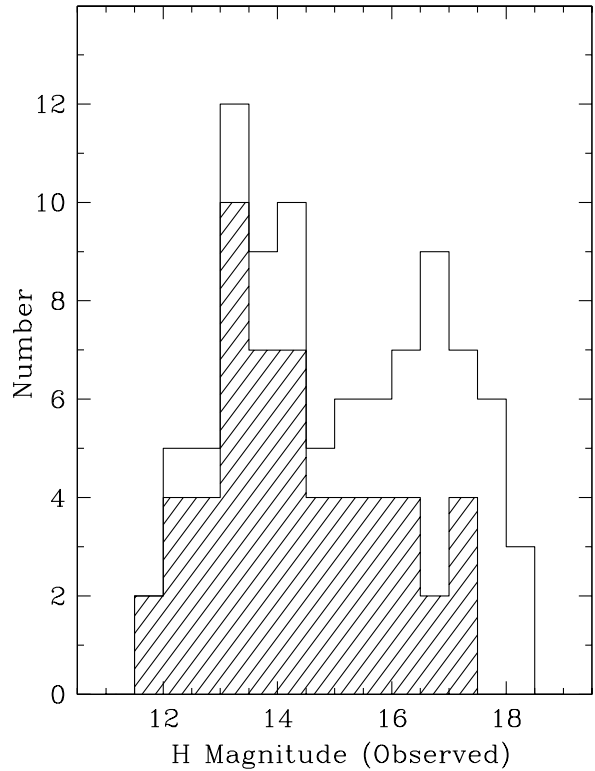}
\caption{\textbf{Histogram of H Magnitudes of Objects Observed.} The shaded area represents objects whose spectra  were classifiable. As expected, the number of objects observed with successful spectroscopy drops at fainter magnitudes, but even at brighter magnitudes it is not completely successful, due to the very different integration times used (and the aperture sizes) on the different runs.}
\label{hist}
\end{figure}

To investigate the completeness of the spectroscopic sample, the observed H magnitudes  of all the objects observed were plotted with respect to whether or not a spectrum was obtained in Figure~\ref{hist}. There is no sharp cutoff, as expected due to the very different aperture sizes and exposure times used for the 4 runs, and the variability in the nebular emission.  The level of completeness is approximately constant for 12 \textless H \textless 16 mag, corresponding to approximately 0.3-0.03 M$_{\sun}$. No objects were classified successfully for H \textgreater 17.5.

\section{Cluster Membership}
\label{membership}

\subsection{Gravity Sensitive Features}
\label{gravity}
Young objects which are still contracting to the MS have surface
gravities intermediate between those of field dwarfs and giants. There
are several optical spectral features, most notably alkali metal
absorption lines, which are extremely sensitive to surface gravity and
hence useful indicators of age and cluster membership.

\begin{figure}
\includegraphics[height=4.5in]{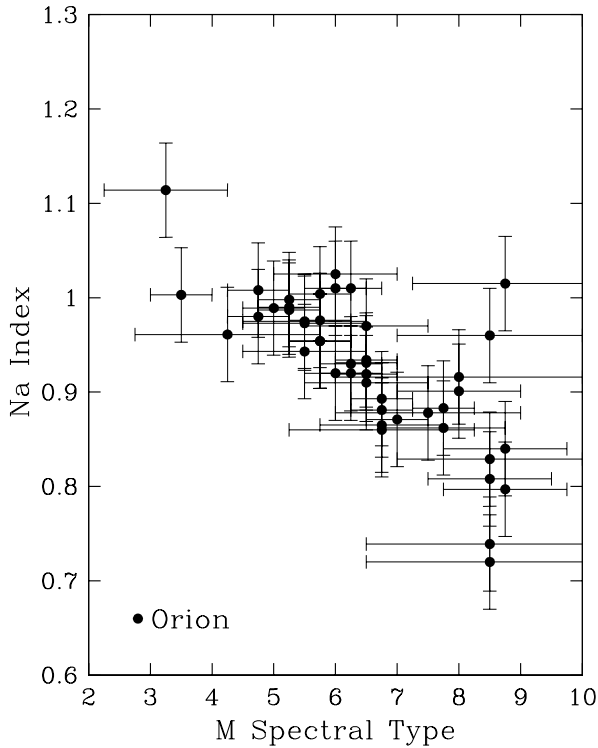}
\caption{Variation of Kirkpatrick et al. (1991) Na Index with spectral type for Trapezium Objects.}  
\label{K91Corion}
\end{figure}

The main form of evidence of cluster membership available in these
spectra is very weak Na I doublet absorption at 8183/8195\,\AA. An
analysis of this feature using the Kirkpatrick et al. (1991) C ratio
was carried out in Paper I for M type objects of different surface
gravities: field dwarfs, giants, dwarf/giant averages and young M type
objects in the Chamaeleon I SFR (having an average age of 2 Myr). We
calculated this ratio for the 45 Trapezium objects with spectral types
assigned -- see Figure~\ref{K91Corion}. 

Studies of the  NaI and KI  absorption lines in the near-infrared show a similar trend, with  
substantially lower equivalent widths in young cluster objects than in field dwarfs (e.g. Allers et al. 2007, Close et al. 2007, Gorlova et al. 2003, Luhman et al. 2007,  LR01)

A deep spectrum of Orion by Osterbrock, Tran \& Veilleux (1992) does not 
reveal any emission lines in the vicinity of the 8183/8195\,\AA\, Na doublet, 
indicating that  the Na C gravity index 
is not contaminated by nebular emission lines.

\vspace{1cm}

The Trapezium spectra have weak Na strengths much
closer to giant than dwarf values, demonstrating their youth and PMS
nature.  The Na index correlates with the spectral type, but the index value for one object, 
183-729 (M8.75), lies above the general trend and almost overlaps with
the values seen for the dwarf spectra. Examination of the spectrum
shows that it does in fact have weak Na absorption and is a low gravity
source and the index was affected by a negative spike in the spectrum that affects the continuum level, making the index higher than it should be.  Source
011-027 (M3.25) has an index value which overlaps that of the
Chamaeleon sequence, this is also very close to the dwarf sequence at
these early M types. Although its Na absorption is closer to the value
for a young object than a dwarf, considering the estimated error
of $\pm$ 0.05 this is inconclusive. The location of this object on the
HRD (Figure~\ref{hrd-fig}) indicates that it is underluminous for its spectral type. Its spectrum 
does not appear to be significantly affected by veiling as we would expect this to also 
weaken the Na absorption.   None of the other objects have
absorption as strong as that seen in the dwarf sequence indicating
that they are all very likely cluster members.

\begin{figure}
\includegraphics[height=4.5in]{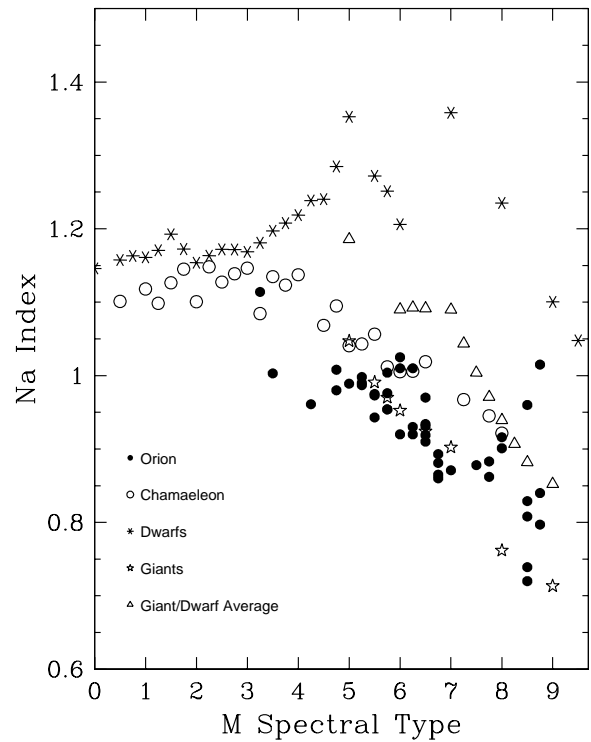}
\caption{Kirkpatrick et al. (1991) Na Index for Trapezium Objects, Chamaeleon I Objects and M Standards. The Na index is plotted against spectral type for objects of different gravity marked by the symbols listed on the figure.  }
\label{K91Call}
\end{figure}

  Figure~\ref{K91Call} compares the Na index values  measured for the Trapezium 
  objects to those of the Chamaeleon objects and the M standards used in Paper 1; 
  the individual Na indices are listed in Table A.1 in the Appendix. 
 The Trapezium objects have slightly lower
 values of the Na index than the Chamaeleon objects, consistently
 lying below them in the plot.  The weaker Na absorption in the
 Trapezium objects indicates that they have lower surface gravity and
 hence are younger on average than the Chamaeleon objects.  However
 the difference is  small (compared with the difference in the index
 calculated for the dwarfs and giants), as expected since the average
 ages of the clusters are similar -- 2 Myr for Chamaeleon (Luhman,
 2004) and 1 Myr for the Trapezium, but does indicate that surface
 gravity  has a relatively strong variation with age at these very
 young ages.  This suggests that careful measurements of the Na index
 may be very helpful in investigating the ages of clusters and perhaps
 even individual objects within clusters.

In order to investigate whether the Na index can provide information
on the ages of the individual Trapezium objects,  the sample was divided into
those lying above and below the 1Myr BCAH isochrone on the H-R diagram
plotted in Figure~\ref{hrd-fig} (see section~\ref{hrd}).  However, no
systematic differences between the Na indices in the two groups were 
found, and it appears that the scatter in
the Na index  is due to random errors incurred with spectral typing,
luminosity estimates and the calculation of the Na index itself,
rather than to systematic age differences within the cluster.

\subsection{Circumstellar Disks}
\label{disks}
A dusty disk surrounding an object is an indication of extreme youth
and hence of cluster  membership. The presence of
circumstellar disks may be inferred in optical spectra by emission from the Ca
\small{II} \normalsize triplet lines at 8498, 8542, 8662\,\AA, which is a
signature of a disk wind.  Circumstellar Ca {\small II} line
emission is detected in several of the Trapezium objects. It is prominent in the 
spectra of 037-246, 017-636, 121-434, and weaker in 030-524, 177-541, 095-058, 
 and 014-413 (and also present in several of the untypable spectra, see 
 Table~\ref{notypes}). This offers additional evidence that these objects are 
 young and hence cluster members.  The equivalent widths of the $\lambda$ 8498\,\AA~ 
 Ca {\small II}  line range from 24\,\AA~ in 121-434 to 3\,\AA~ in 014-413;  the highest value, 
 36\,\AA, is in the untypable object 047-436 where a combination of vigorous accretion 
 and extinction presumably combine to obscure the stellar absorption features.  The  
 spread in the spectral types given by the indices is no greater in the objects 
 with  Ca {\small II}  triplet emission  than the objects where  Ca {\small II} 
 emission is not detected (see Table A.1), suggesting that contamination by continuum associated 
 with accretion does not have a significant effect on the spectral classification; this is confirmed by visible inspection and comparison with template spectra.  Note that 177-541 also displays prominent emission lines at 8446, 8578 and 8729\,\AA\, which are much brighter than the Ca {\small II} triplet emission; the lines presumably arise from [O I], [Cl II] and [C I]/HeI. The [O I] and [Cl II] lines are  present in the deep Orion spectrum presented by Osterbrock, Tran \& Veilleux  (1992), but are much brighter than the nearby Paschen H lines in 177-541, which suggests that they are associated with the star rather than just being due to residual nebular emission.

\subsection{Membership Statistics}
Hillenbrand \& Carpenter (2000) calculated the expected foreground and background contamination in the central 5.1x5.1 square arcmin using a modified version of the Galactic star count model of Wainscoat et al. (1992) and taking into account the variable extinction across the region. This, in combination with the photometric study of Muench et al. (2002), indicates that contamination by background objects starts to become significant at K = 16, rising at fainter magnitudes and perhaps peaking at K = 19-20 (Hillenbrand \& Carpenter 2000). This magnitude limit includes all the objects observed here with spectral types assigned, therefore background contamination is likely negligible, and we would expect less than one field star in the dataset of 45 spectra.

\section{Construction of the HRD}
\label{hrd}
In order to place an object on the HRD, its temperature must first be established, as well as either its bolometric luminosity or magnitude in a single band.  If the diagram is overlaid with isochrones then the age spread of the cluster can be determined. An average age for the cluster can then be applied to all members in order to estimate their masses. Both the spectral typing and the conversion to $T_{\rm eff}$ are critical since a star's mass and age derived from its position on an H-R diagram is primarily a function of its assigned $T_{\rm eff}$, i.e. its observed spectral type, since at these ages the objects lie on primarily vertical tracks (e.g. BCAH98; see Figure~\ref{hrd-fig}).  Hence uncertainties in the masses derive mainly from uncertainties in the spectral types. 
 
\subsection{Dwarf, Giant and Intermediate Temperature Scales}
\label{temp}
Since evolutionary models are given in terms of $T_{\rm eff}$ and not spectral type, a conversion is required from the spectral types obtained from the spectra. The main difficulty with estimating ages and masses for PMS stars from the H-R diagram is the establishment of a reliable temperature scale for young T-Tauri like objects.  The temperature scale for late-M dwarfs is highly uncertain (Luhman \& Rieke 1998; Gorlova et al. 2003), and in particular the same spectral type will not correspond to the same temperature in a dwarf, PMS object or giant, so temperature scales have to be defined for all these luminosity classes.  

For G and K stars, giants are cooler than dwarfs whereas M giants are warmer than M dwarfs for the same spectral type, the crossover being at M0 (Luhman 1999), although the magnitude of the temperature difference varies with spectral type and the scale used.  The M giant scale of Luhman (1999) is warmer than the M dwarf scale by 200-400K, the difference being largest at mid-M types, where most of the data in this study lie.  Hillenbrand \& White (2004) find even larger differences: M2, M4 and M6 types being 310, 500 and 620K warmer for giants than dwarfs respectively. For M stars, the adoption of an intermediate temperature scale instead of one for dwarfs therefore moves the stars to older ages and higher masses on the evolutionary tracks than those obtained from using dwarf temperature scales (and more so at lower masses). 

It is therefore vital to use a suitable scale for conversion and so, when choosing an appropriate temperature scale, the expected age of the objects must be taken into account.  With the expected average age of the Trapezium objects being $\sim$ 1 Myr, a dwarf scale is inappropriate due to the lower gravity of PMS stars.

\subsubsection{Temperature Conversion}
\label{tscale}
We have used the scale of Luhman et al. (2003b) to convert from spectral types to $T_{\rm eff}$.  The scale is an extension to M9 of that in Luhman (1999), which begins at M0 and ends at M7.5, using data from Taurus and IC348, following the trend for earlier types and assigning temperatures intermediate between dwarf and giant scales.  This scale has been used extensively for young objects, e.g. in the Chamaeleon I region by Luhman (2004) and Neuhauser \& Comeron (1999) and in Taurus by White \& Basri (2003).  This scale is also compatible with BCAH98 models.  

The temperature difference between M8 and M9 is larger than between other subtypes. The low temperature derived for types later than M8 therefore has the effect of moving these objects to lower masses and younger ages on the HRD if this scale is correct.  Note that the position of the latest spectral types on the H-R diagram is very sensitive to the temperature scale adopted. 



\subsection{Luminosity}
\label{lum}
Since mass estimates for young low mass stars are relatively insensitive to uncertainties in the luminosity (except at ages younger than 1 Myr),  the choice of band and the calculation of luminosity is not as critical as the spectral typing and temperature conversion for the estimation of masses, although luminosity errors will affect the derived ages to a greater extent.

Synthetic colours available in evolutionary models such as those of BCAH98 provide a more accurate determination of the intrinsic properties from the observed magnitude and/or colour. These evolutionary tracks provide self-consistent flux predictions, permitting luminosity and mass to be obtained without the need for empirical bolometric corrections. This is especially advantageous for stars younger than 10 Myr since they are embedded in molecular clouds and often have excess UV (accretion) and IR (disk) emission, so the bolometric stellar luminosity cannot be derived by simply integrating the broadband photometry. To avoid contamination by these types of nonstellar emission, the luminosity must be derived using measurements at wavelengths where veiling is expected to be relatively small. 

\begin{figure}
\includegraphics[height=4.5in]{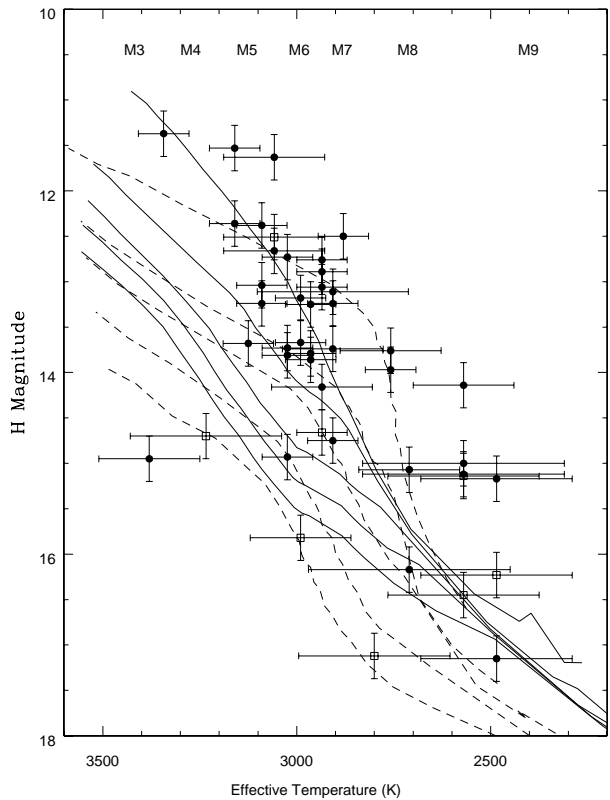}
\caption{The Trapezium Cluster HR.D The isochrones of BCAH98 are shown for ages of 1, 3, 5, 7 and 10 Myr (solid lines, top to bottom) and those of DM97 are shown for ages of 1, 3, 5, 10 and 20 Myr (dashed lines, top to bottom).  Objects which have anomalously blue colours are plotted as open squares.  The origin of the error bars is described in the text.  The dereddened H-magnitudes of the 45 objects which have been successfully typed are plotted on the  left hand ordinate, while the total luminosity is shown on the right hand side after application of bolometric corrections.  }
\label{hrd-fig}
\end{figure}

To compare the observations with models, we plot $T_{\rm eff}$ against the dereddened H magnitudes obtained from the photometry of LR00 and Lucas et al. (2005).  Figure~\ref{hrd-fig} shows the HRD for the Trapezium objects, plotted with the isochrones of BCAH98 and D'Antona \& Mazzitelli (1997). The dereddened H magnitudes are converted to solar luminosities by the application of a bolometric correction which takes the form Log(L$_\odot$) = -0.401 + 4.14(H$_{\rm dered}$) in this luminosity range at the adopted distance of 450 pc; this scale is plotted on the right hand ordinate.


\subsection{Sources of Error on the HRD}

For the majority of sources earlier than M8 the uncertainty in the spectral type is less than $\pm$ 0.5 subtypes, as shown in Table~\ref{types}. The uncertainty in temperature derived from the spectral type therefore consists of this random error in addition to any systematic error present in the temperature scale itself,  e.g. due to the surface gravity not being exactly that of the average of dwarfs and giants (as discussed in Section~\ref{tscale}).

Errors in the luminosity arise from errors in the photometry, which are typically 0.1 mag -- see 
LR00 and Lucas et al. (2005) for a detailed discussion. We have conservatively assumed a constant 
0.25 mag error to account for additional errors due to dereddening and scattered light effects in these young objects. Variability may also be present and could be quite significant;  for example, Scholz \& Eisl\"{o}ffel (2004) find several LMS and BDs in $\epsilon$-Ori show variability attributed to rotation, accretion and flares. The spectral variability of these Trapezium sources is investigated via a comparison of the spectra between the different runs in Section~\ref{duplicates} and found to be unobservable within the estimated spectral typing errors.

\subsection{Evolutionary Models}
\label{tracks}

We use the Lyon Group models for the analysis of these data. The models provide mass-age-colour-magnitude relationships and are compatible with the temperature scale used. The 'NextGen' (BCAH98)  models we use are appropriate for dust-free atmospheres at T$_{\rm eff}$ $>$ 2400K and are calculated using nongrey model atmospheres. There are no objects in this sample with $T_{\rm eff}$ $<$ 2400K and so dust formation is not likely to affect the spectra significantly.  This is supported by the strength of the TiO and VO molecular bands present in all the spectra and in the values of the indices calculated (see Paper I) -- they are strong in all the spectra and have not started decreasing in strength at types earlier than M9, as would be expected  if dust condensation occurred.

Other models have also been produced and used widely, e.g. Burrows et al (1997; BM97) and D'Antona \& Mazzitelli (1997, updated in 1998; DM97), as used by LR01 for the estimation of masses for many of the same objects.

\section{Analysis of the HRD}
\label{analysis}
\subsection{Ages \& Star-Formation History}
\label{hrd-ages}

The technique of estimating ages and masses from the HRD by adopting
an average age for the cluster has been used in a number of star formation regions, e.g. in Taurus
(Brice\~no et al. (2002),  White \& Basri (2003)),  in Rho-Ophiuchus
(Wilking et al. 1999) in NGC 2024 (Levine et al 2006), and in Orion  (Slesnick et al. 2004).

Age estimates become highly uncertain at very young ages, because the
state of evolution depends critically on the initial conditions.   The
BCAH98 models do not  cover ages younger than 1 Myr

Different ages have been estimated for Orion. Hillenbrand (1997) found
that the ONC as a whole has a mean age of 0.8 Myr and an age spread
of less than 2 Myr, with most SF occurring 0.3-2 Myr ago with a small
tail out to 10~Myr. DM97 tracks show that 50\% of the SF has occurred
in the last 0.5 Myr and 85\% within the last 2 Myr, therefore
adopting an average of 1 Myr seems  appropriate when estimating
masses. However, there is also evidence that previous episodes of SF have occurred,
e.g. Slesnick et al. (2004) find a distinct older population
at $\sim$10~Myr. Palla \& Stahler (1999) also found a low level of SF
10~Myr ago, after which it accelerated to the present epoch and that
the best fit age is 2~Myr. Luhman et al. (2000) find a median age for
the Trapezium of ~0.4~Myr for spectral types  as late as M6 using
DM97 models.  For a discussion of sources of error and the associated problems 
of disentangling errors in observation and theoretical calculations from true
 variations in age and age spread within young clusters, see 
 Hillenbrand, Bauermeister \& White (2007).   

Palla et al. (2005) studied a sample of 84 low-mass stars in the range
~0.4-1 M$_{\sun}$, and with isochronal ages greater than $\sim$ 1 Myr
from the Hillenbrand (1997) survey with membership probability greater
than 90\% to investigate if sub-solar members have had time to
significantly deplete their initial Li content. For the subsample with
M $<$ 0.6 M$_{\sun}$, in the case of stars with isochronal ages $>$ 3
Myr, 4 show significantly depleted Li abundances. The derived ages of
$\sim$10~Myr indicate that the ONC does contain objects older than
the average age of the dominant population, indicating that SF
duration extends into the past, although at a reduced rate. The
presence of a population with isochronal ages in excess of $\sim$10~Myr 
has also been inferred by Slesnick et al. (2004) in their study of
the very low mass stars and brown dwarf members. However, this
apparently old population contains at least some objects with
significant contamination from scattered light, which could lead to
overestimation of the temperature and underestimation of the true 
luminosity (see below). 

The 45 objects with identified spectral types are plotted on the HRD
in Figure~\ref{hrd-fig} along with isochrones from BCAH and DM97.
The two sets of isochrones differ significantly in slope at late spectral types ($>$M6)
with the DM97 isochrones being significantly steeper than BCAH.  
23 of the 45 objects lie above the 1~Myr BCAH isochrone, while only 7 lie below the 5~Myr isochrone.  This large fraction lying above the youngest isochrone implies that there are
many very young objects, but their ages are not estimable from these
tracks.  The 1 Myr DM97 track passes through the region populated   by objects with spectral types between M7 and M8 and might suggest a slightly older population, 
but again 19 objects lie above the 1~Myr isochrone, and only 6 objects 
below the 5~Myr isochrone.    The fluxes were checked by constructing a 
similar diagram using the non-contemporaneous JHK photometry from Lucas et al.(2005) 
(where available). The results were found to be very similar, but with a 
slight tendency for sources to shift to higher luminosity and even younger 
ages (but by only 0.1 mag on average).

Objects earlier than M7 scatter about the 1 Myr BCAH isochrone, consistent
with the many other estimates of the mean age of the cluster around 1
Myr, but the majority of the objects with types later than M7 lie
above this isochrone.  If the DM97 isochrones are adopted, the objects 
earlier than M7 tend to lie below the 1Myr isochrone, but the later spectral types tend to lie above it. 
However, there are very significant selection effects in our sample, and especially at spectral types beyond M7. Our sample was selected to lie primarily within the 12.5 $<$ H $<$ 15 magnitude range which should give an unbiased distribution of spectral types within that range, but at each
spectral type there may be a bias towards bright or faint objects
depending on the location of the isochrones within that range. The
fact that we detect objects primarily above the isochrone is at least
partially because we will tend to obtain better quality spectra of the brighter objects at a
given spectral type, biasing the distribution. At types later than M7
we are not sensitive to the expected population of faint low mass BDs
at ages of 1 Myr or more,  and most objects detected are located
above the isochrones with apparent ages less than 1 Myr.  Similar effects 
are seen by Levine et al (2006) in the NGC 2024 cluster.  At early M
types, the bulk of the stellar population is brighter than the
magnitude range of our sample, so we would expect to detect 
fainter, older objects of stellar mass (if they exist).  

Infrared spectroscopy  in Orion has confirmed the existence of  a 
population of faint (H$>$ 17.5 mag) objects  with deep water absorption bands which 
confirm them as very late M-type or early L-type low luminosity brown dwarfs or planetary 
mass objects(Lucas et al 2006).  These lie below the 1 Myr isochrone indicating that a  
bias towards brighter objects operates on the sample discussed in this paper.  

The other possible explanation, that the least massive members tend to
be younger than more massive members, is unlikely to be the cause. There
could however be an effect from the unreliability of isochrones at such young
ages caused by the lack of a stellar birthline in the evolutionary
calculations, which could lead to the models underpredicting the luminosity of very
young low-mass objects. Alternatively, some of the objects located
well above the isochrone could be binaries with nearly identical
components.

A few apparently older objects, lying below the 5 Myr isochrones are also seen, but once possible
spectral typing and luminosity uncertainties have been taken into account, 
there remain just one or two objects likely to be older than 5 Myr.  
The depth of the Na absorption in 011-027 places it close to the Chamaeleon
values, but within the errors, it could also be located on the M dwarf
sequence. It is therefore not conclusively either a cluster member or 
nonmember.  Four other objects  lie below the 10~Myr BCAH isochrone, but three of them, 177-541, 
156-547 and 030-524,  have anomalously blue (I-J) colours, which 
are consistent with either a large age or an exceptionally young age (see below).  
177-541 is located in the brightest part of the Orion bar, hence the 
error bars for its flux and temperature are probably underestimated, and as discussed in section 4.2, it has the spectral signatures of strong accretion.

\begin{figure}
\includegraphics[height=4.5in]{figs/hrd_sles2.epsi}
\caption{HRD for the data from this paper  (filled symbols) together with those of Slesnick et al. (2004) (open circles). 
The solid lines represent the BCAH 1, 3, 5, 7 and 10~Myr isochrones while the dashed lines are the mass tracks for 0.5, 0.4. 0.3, 0.2, 0.15, 0.1, 0.08, 0.07, 0.06, 0.04, 0.03, 0.02 and 0.15 M$_\odot$ from left to right.}
\label{hrd-slesnick}
\end{figure}

\subsection{Masses}
\label{hrd-masses}
The masses derived for objects on the H-R diagram are relatively insensitive to the age adopted, and are more sensitive to the spectral type or temperature assigned.  This is because the tracks at young ages are primarily vertical (see Figure~\ref{hrd-slesnick}) so that an age uncertainty of a few Myr  will lead to a relatively small uncertainty in the derived masses. The objects with spectral types in the range M4 to M7, where the instrument sensitivity permits good sampling of a wide range of ages, scatter about the 1 Myr isochrone. For this reason, and consistent with other studies cited above, we adopt an average age of 1~Myr.  

Table~\ref{results} gives the masses found using BCAH98 tracks and an assumed age of 1 Myr. These masses are also compared where possible with the masses from photometry of LR00 and Lucas et al. (2005) for objects in common.

\begin{figure*}
      \centering
   \resizebox{\hsize}{!}{\includegraphics[clip=true]{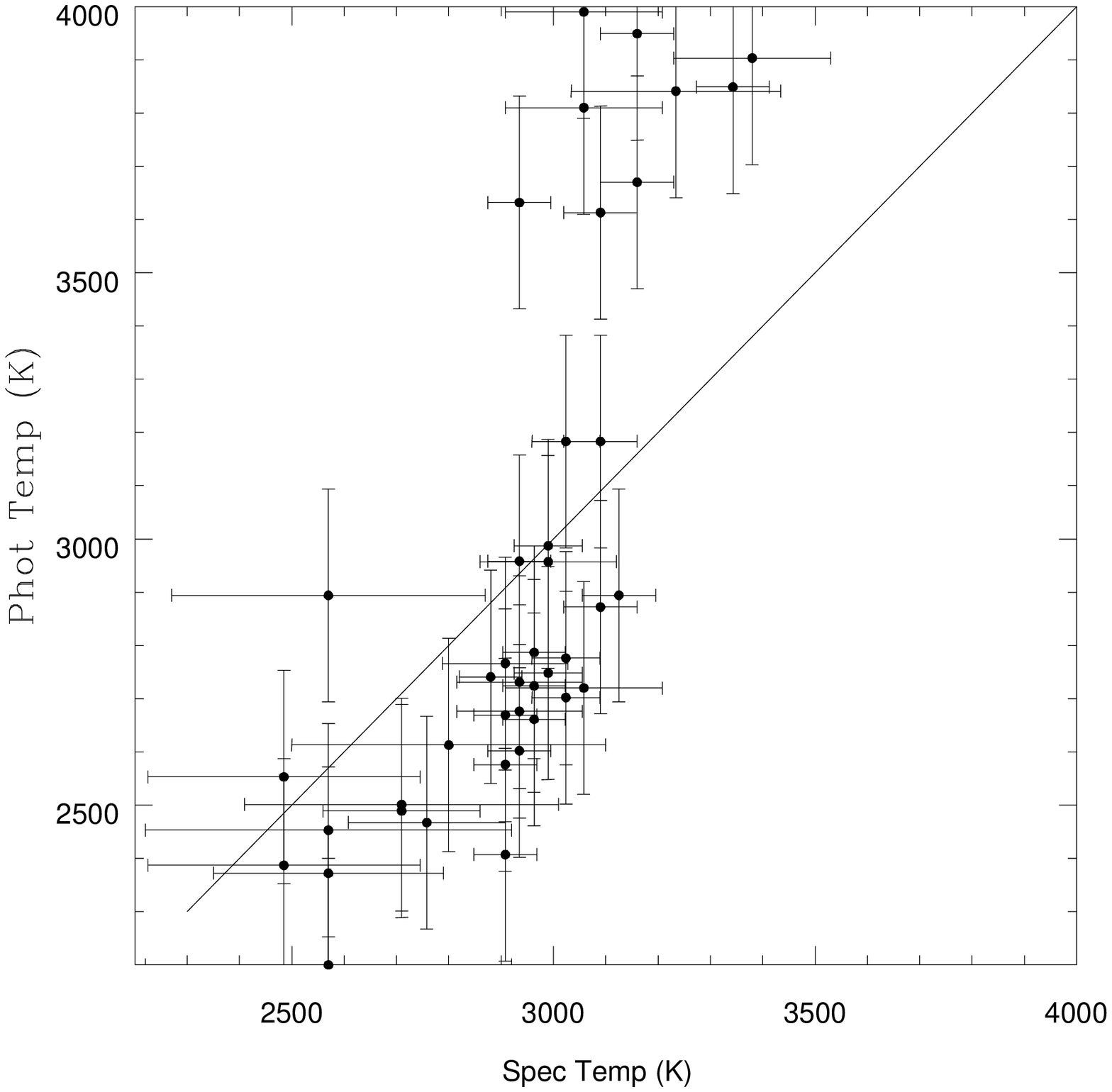} 
  \includegraphics[clip=true]{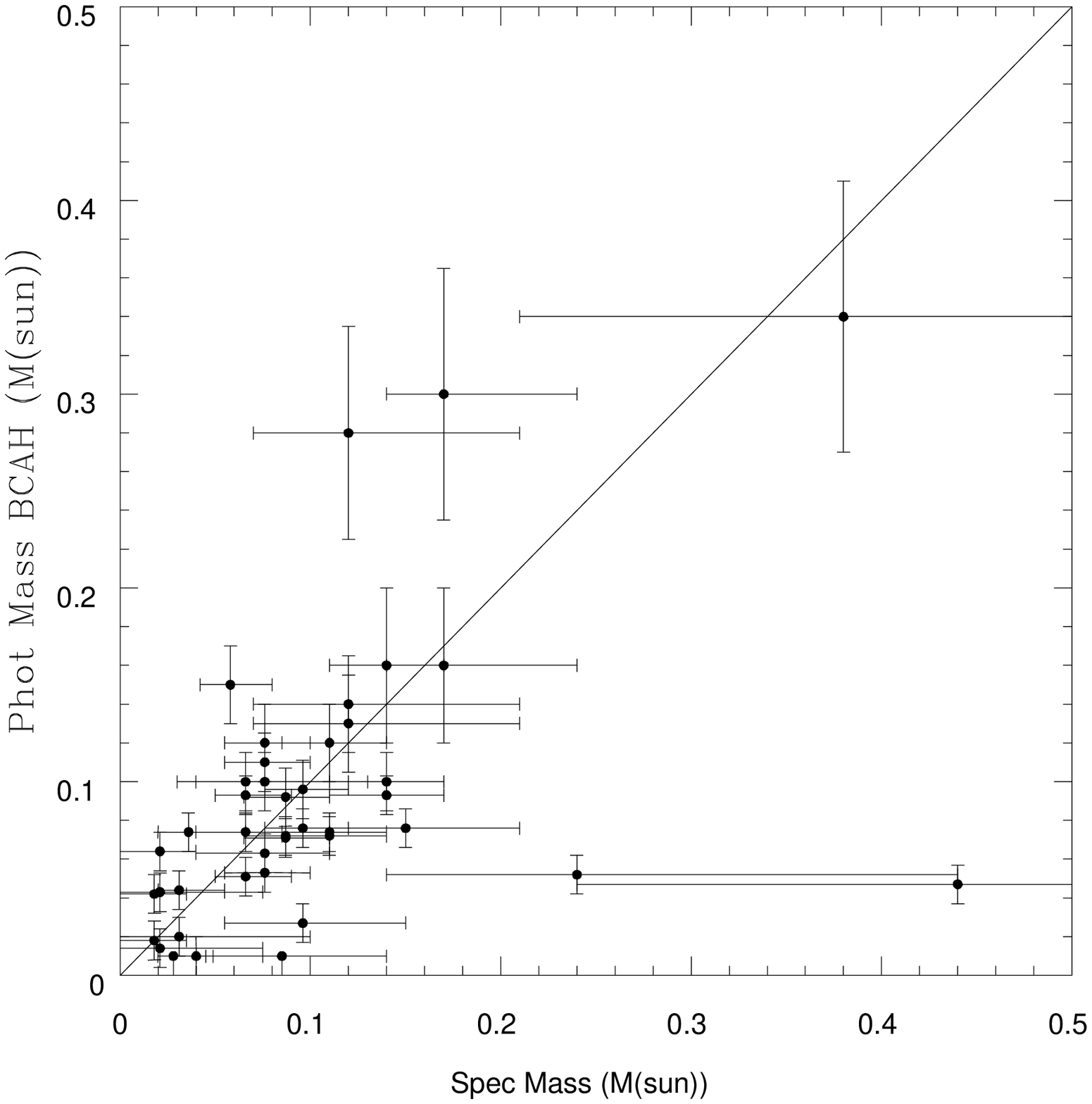}
   }

  \vspace*{-33mm}
     \caption{ Left:  Comparison of photometric temperatures, derived from dereddened (I-J) photometry, with the spectroscopic temperatures obtained in this work.    Right: Comparison of photometric masses derived from the dereddened H-band fluxes and the BCAH 1~Myr isochrone with the masses derived from the spectral types. }
        \label{tempcomp}
    \end{figure*}

\subsection{Comparison with HRD of Slesnick et al (2004)}
\label{slesnick}
Slesnick et al. (2004) constructed an HRD based on near-infrared spectroscopy for $\sim$100 M-type LMS and BDs in the inner 5.1x5.1 arcmin of the ONC and found two distinct groups of objects: a young population with an average age of less than 1 Myr and a significant older population with an average age of $\sim$10 Myr.  The apparently older population constitutes a relatively uniform population across the mass range.  

A similar situation is seen in Taurus where Luhman et al. (2003a) found that some sources appeared anomalously old. In contrast to these results, Comeron et al. (1999) found that most of the stars in Cha I are of similar age and no significant older population is seen. Comeron et al. (2000) also found the objects to be highly clustered on the 2 Myr isochrone, suggesting an essentially coeval population having formed over a timespan of less than 1 Myr.  Similarly in IC348 (1-2 Myr) and Rho-Oph (0.4 Myr), the data are consistent with very short durations for the SF (Luhman et al. 2000).

We have plotted the Slesnick et al  (2004) data points together with the spectral types obtained in the current work  on the HRD in Figure~\ref{hrd-slesnick}, using the same temperature scale and tracks as Figure 4.  To enable direct comparison, we have calculated the dereddened H magnitudes from the data (A$_{\rm V}$ and IR photometry) presented by Slesnick et al, and used the spectral types assigned in that paper rather than the temperatures, as the temperature scale used is quite different from that in the current work.  The paucity of objects older than 5 Myr in our data is in contrast to the Slesnick et al. (2004) finding of a distinct population of older objects with average ages of 10 Myr.   Most of the objects in our sample are located further from the cluster centre and the dense molecular ridge of OMC-1 than the sample of Slesnick et al., so one might have expected to find older ages than in that study. The
apparently old sources in the Slesnick et al. sample include only 1 or 2 objects with higher 
reddening than our sample (which was selected for low extinction) so the old population would
not have been excluded from this study by extinction.

Slesnick et al (2004)  pointed out that one of the objects in the group with ages of $\sim$10~Myr is a proplyd and a further two coincide with silhouette disks.   Inspection of Hubble Space Telescope (HST) H$\alpha$ data for Orion (see O'Dell \& Wong 1996 and the larger Treasury dataset of Robberto et al.2004) may resolve the discrepancy  for at least a few more  of the apparently old sources. 
We examined the Treasury data for the 14 'old' sources from Slesnick et 
al. (2004) (defined by spectral type $<$M4 and K mag$>$13) and found that 4 of them (29\%) are 
spatially resolved with disks (3 are included in the list of O'Dell \& Wong 1996). 
These are very young ($\ll$1~Myr) stars surrounded by the remnants of
their natal cloud cores, which are externally illuminated by the O-type stars in the centre of
the cluster. We deduce that the photospheres of such sources are very strongly veiled by a 
combination of scattered light from the O-type stars, scattered light from the source itself
and perhaps free-free emission from a stellar wind in these strongly accreting objects.
This veiling would lead to overestimation of the temperature and (via an underestimate
of the extinction) an underestimation of the luminosity). Furthermore, a circumstellar disk may occult the flux from the photosphere, reducing the apparent luminosity of such objects (e.g. Luhman et al 2003b). The 4 proplyd sources are numbers 25, 30, 127 and 143 from Slesnick et al.  Number 127 is the well-known silhouette disc source 114-426 (McCaughrean et al.1998). 

It is possible that some of the other 10 apparently old sources in the Slesnick et al. list 
are also strongly veiled protostars (aged $\ll$1~Myr) which receive insufficient ultraviolet 
flux  from the O-type stars to be detected in H$\alpha$ imaging. In this context it is interesting 
to note that of the 5 ``old'' sources detected at I band in the imaging dataset of LR00, all have 
anomalously blue $(I-J)$ colours for their magnitudes, as defined in that work. These blue colours 
were attributed to scattered light, owing to an almost one to one correspondence between 
proplyd classification from the HST H$\alpha$ imaging and blue colour. Three of the proplyds
mentioned above have I band detections and all of these have blue colours. However blue $(I-J)$ 
colours could also be explained by a warm photosphere so it remains possible that some of these 
sources genuinely have ages of $\sim10$~Myr.

\begin{table*}

\begin{center}
\caption{\textbf{ Effective Temperatures, Masses and Luminosities of Trapezium Objects from Spectroscopy or Photometry.} }
\label{results}
\begin{tabular*}{420pt}{cllclcclll}
\hline
Name & Spectral &  $T_{\rm eff} $ & $T_{\rm eff}$  & Mass & Low M & Upp M &M  ($M_{\sun}$)  &M ($M_{\sun}$) & Log($L$)  \\
 & Type	 & (K) & (K) & ($M_{\sun}$)  &  ($M_{\sun}$) & ($M_{\sun}$)  & (DM97) & (BCAH98)  & 
($L_{\sun}$)\\
& Assigned & Spect$^a$ & Phot$^b$. & Spect$^c$. & & & Phot$^d$  & Phot$^e$ & Phot$^f$\\
\hline
011-027 &M3.25&3380$\pm$ 150& 3903&0.44&0.24&0.62&0.048&0.047 & -1.86\\
4584-117 &M3.5&3343$\pm$ 70& 3849  &0.38&0.21&0.57&0.33 & 0.34	& -0.42 \\
177-541 &M4.25&3234$\pm$ 200& 3841&0.24 & 0.14&0.44&0.054 &0.052	& -1.76 \\
112-532 &M4.75&3160 $\pm$ 70& 3670 &0.17&0.14&0.24&0.18 &0.16 	& -0.82 \\ 
082-403 &M4.75&3160$\pm$ 70& 3949 &0.17&0.14&0.24&0.30 &0.30 & -0.49 \\
091-017 &M5.0 &3125$\pm$  70 & 2894 &0.15&0.12&0.21&0.087 & 0.076 & -1.35 \\
034-610 &M5.25&3090$\pm$ 70 & 3183 &0.14&0.11&0.17&0.11 & 0.093 & -1.17 \\
016-534 &M5.25&3090$\pm$ 70 & 2872 &0.14&0.11&0.17&0.12&0.10	& -1.09 \\
130-458 &M5.25&3090$\pm$ 70 & 3613 &0.14&0.11&0.17&0.18&0.16	& -0.83 \\
121-434 &M5.5&3058$\pm$  150& 3990 &0.12&0.07&0.21&0.15 & 0.13 	& -0.94 \\
017-636 &M5.5&3058$\pm$ 150 & 3810 &0.12&0.07&0.21&0.17 &0.14	& -0.88 \\
037-246 &M5.5&3058$\pm$ 150 & 2720 &0.12&0.07&0.21&0.29&0.28	& -0.53 \\
222-745 &M5.75&3024$\pm$ 65 &  &0.11&0.085&0.14& 0.049 &0.047	& -1.85 \\
068-019 &M5.75&3024$\pm$ 65 & 2776 &0.11&0.085&0.14&0.085 & 0.074  & -1.37\\
019-354 &M5.75&3024$\pm$ 65 & 2702 &0.11&0.085&0.14&0.081&0.072 	& -1.40 \\
103-157 &M5.75&3024$\pm$ 65 & 3183 &0.11&0.085&0.14&0.15 &0.12 	& -0.97 \\
017-710 &M6.0&2990$\pm$ 65 & 2748 &0.096&0.076&0.12& 0.11 &0.096	& -1.15 \\
069-209 &M6.0&2990$\pm$ 65 & 2987 &0.096&0.076&0.12&0.087&0.076	& -1.34 \\
156-547 &M6.0&2990$\pm$ 130 & 2957 &0.096&0.055&0.15& 0.033 &0.027 	& -2.21 \\
4559-109 &M6.25&2963$\pm$ 60 & 2787 &0.087&0.065&0.11&0.082 & 0.072 & -1.39 \\
102-102 &M6.25&2963$\pm$ 60 & 2724 &0.087&0.065&0.11&0.079 & 0.071 	& -1.42 \\
095-058 &M6.25&2963$\pm$ 60 & 2661 &0.087&0.065&0.11&0.11 & 0.092	& -1.18 \\
077-453 &M6.5&2935$\pm$ 60 & 3632 &0.076&0.055&0.10 & 0.14 &0.12 	& -0.98 \\
072-638 &M6.5&2935$\pm$ 60 & 2958 &0.076&0.055&0.10 &0.055 &0.053	& -1.74 \\
053-503 &M6.5&2935$\pm$ 120 & 2731 &0.076&0.040 &0.12&0.12 &0.10 	& -1.10 \\
154-600 &M6.5&2935$\pm$ 60 & 2602 &0.076&0.055 &0.10 & 0.13 &0.11 	& -1.03 \\
014-413 &M6.5&2935$\pm$ 120 & 2676 &0.076&0.040 &0.11  &0.069&0.063 & -1.54 \\
4569-122 &M6.75&2908$\pm$ 120 & 2766 &0.066&0.040 &0.11 &0.084 & 0.074 & -1.37 \\
035-333 &M6.75&2908$\pm$ 60 & 2576 &0.066&0.05 &0.09 &0.11 &0.093 	& -1.17 \\
096-1943 &M6.75&2908$\pm$ 60 & 2669 &0.066&0.05 &0.09 &0.053 & 0.051 & -1.78 \\
055-230 &M6.75&2908$\pm$ 180 & 2407 &0.066&0.03 &0.13 &0.12&0.10 	& -1.12 \\
084-305 &M7.0&2880$\pm$ 60  & 2741 &0.058&0.042 &0.08 &0.17 &0.15 	& -0.87 \\
030-524 &M7.5&2800$\pm$ 300 &2613 & 0.040 &  0.02 & 0.14 & 0.025 & 0.010 & -2.73 \\
217-653 &M7.75&2758$\pm$ 80 &  &0.036&0.028 &0.045 & 0.075 &0.068	& -1.46 \\
042-012 &M7.75&2758$\pm$ 150  & 2467 &0.036& 0.020 & 0.066 &0.083 & 0.074 & -1.38 \\
092-606 &M8.0&2710$\pm$ 300 & 2501 &0.031&$<$0.014 &0.10& 0.029 &0.020 & -2.35\\
186-631 &M8.0&2710$\pm$ 150 & 2489 &0.031& 0.020 & 0.055 & 0.046 &0.044  & -1.90 \\
130-053 &M8.5&2570$\pm$ 350  & $<$2200 &0.021& $<$0.014& 0.075 &0.045 & 0.043 & -1.92 \\
047-550 &M8.5&2570$\pm$ 350  & 2453 &0.021&$<$0.014 &0.075 & 0.026 &0.014 & -2.46\\
077-127 &M8.5&2570$\pm$ 220 & 2372 &0.021&$<$0.014 &0.040&0.069 & 0.064 & -1.53\\
082-253 &M8.5&2570$\pm$ 300  & 2894 & 0.021 &  $<$0.014 & 0.055 & 0.044 & 0.043  & -1.93\\
148-831 &M8.5&2570$\pm$ 350 &  &0.021&$<$0.014 &0.070 & 0.047 & 0.046 & -1.88\\
183-729 &M8.75&2485$\pm$ 350 &  &0.018&$<$0.014&0.045 & 0.026 &0.010	& -2.74 \\
165-634 &M8.75&2485$\pm$ 260 & 2553 &0.018&$<$0.014&0.035&0.044 &0.042 & -1.95\\
031-536 &M8.75&2485$\pm$ 260 & 2387 &0.018&$<$0.014&0.035& 0.028 &0.018 & -2.37 \\

\hline
\end{tabular*}
\end{center}

\begin{tablenotes}
\item Notes:
\item a) The uncertainties on the spectroscopic effective temperatures are representative errors based on the spectral type error only and do not include systematic errors in the temperature scale. At late spectral types, the quoted errors become asymmetric with the error on the low temperature side increasing to ~1.5 times the value quoted.
\item b) Photometric temperatures estimated from the dereddened infrared photometry. Typical uncertainties are $\pm$200 K (see text)
\item c) Mass derived from spectroscopic temperatures and the 1 Myr BCAH isochrone.  Upper and lower mass bounds are derived from the errors in the temperatures. 
\item d) Mass estimated from the dereddened H-band fluxes  and the 1 Myr DM97 isochrone
\item e) Mass estimated from the dereddened H-band  fluxes and the 1 Myr BCAH isochrone
\item f)  Logarithm of the Bolometric Luminosity derived from the dereddened H-band fluxes (see section 5.2); the photometric uncertainties of 0.25 mag. give uncertainties of 0.1 dex in Log($L$).   
\end{tablenotes} 
\end{table*}

\subsection{Spectral Types and Masses from IR Data}

Many of the objects in this sample have also been observed by LR00,
LR01, and Lucas et al. (2005) using broad-band photometry or  H and K
spectroscopy. From the spectra, they defined  IR water indices and
compare these to the  dusty model spectra of Allard et al. (2000;
2001) to derive $T_{\rm eff}$ and obtain spectral types from a comparison
of the indices with those of local field dwarfs. Temperatures were
calculated from the dereddened photometric $(I-J)$ colour using the Nextgen 
model prediction for age 1~Myr.  The
uncertainties in the temperatures derived from photometry are 
typically $\pm 200$~K (see LR00) (not including systematic 
uncertainties in the T$_{eff}$ vs. $(I-J)$ relation) but may be larger 
in the case of veiled proplyd sources.  Where no photometric temperature
is given, this was unobtainable due to the lack of an I band
magnitude.  The temperatures and masses derived  from the infrared fluxes are 
compared with our new spectroscopic estimates in Table~\ref{results}. Spectral
typing errors are given in Table~2 and from these the
spectroscopic temperature errors are calculated individually. 
Whilst a trend is present in the data in the sense that objects with higher temperatures 
assigned from spectroscopy are also hotter from the photometric measurements, the 
agreement is poor (see Figure~\ref{tempcomp}a) with the photometric measurements giving much higher temperatures for the earliest spectral types and cooler temperatures for the later types. We attribute this to errors in de-reddening, contamination by scattered light and photometric uncertainties, and believe the spectroscopically-derived temperatures  to be much more reliable.  

Photometric masses were derived from the 
luminosity vs mass relations at 1~Myr as predicted by the BCAH 
and the  DM97 isochrones.  The masses are estimated simply by matching 
the dereddened H-band magnitudes to the fluxes predicted by the isochrones 
at the distance of Orion; as discussed in section~\ref{hrd-masses}, the mass is 
relatively insensitive to the adopted age. 
The photometric masses derived from the DM97 and BCAH isochrones are quite similar, 
but with the former tending to give somewhat higher masses than the latter.  
Spectroscopic masses use the BCAH98 models assuming an age
of 1 Myr. The lower and upper mass limits are calculated from the estimated 
spectral typing errors for each object, as listed in Table~2.   In some cases, 
the latest possible spectral type is later than M9 where the temperature
scale is not defined, so a lower limit to the mass of 0.014 $M_{\sun}$ is quoted.   
The spectroscopic masses are compared with the photometric masses derived 
from the  BCAH isochrone in Figure~\ref{tempcomp}b.  There is agreement overall 
between the two estimates, but with a number of outlying objects.  The objects with 
significantly lower photometric masses than spectroscopic masses are those that 
lie below the 1~Myr isochrone in Fig~\ref{hrd-fig} and, as discussed above, many have anomalous photometric colours.  

The range of spectral types for this dataset is M3.25-M8.75 and at 1
Myr using the BCAH98 models these correspond to masses ranging from
0.018-0.44 $M_{\sun}$, i.e. from brown dwarfs just above the deuterium-burning mass limit 
to low-mass stars well above the hydrogen-burning mass limit. 
Agreement between the masses obtained from these optical spectra
and those from the infrared photometric studies is variable and
similar to that found by Luhman et al. (1998b) and Slesnick et al (2004) where IR types differ
by up to 3 subtypes  from optical types. However, for most objects, the photometric and spectroscopic mass determinations are broadly consistent.

\subsection{Binarity and its Effects on Spectroscopy \& Photometry}
\label{binarity}

In this study, we do not make any allowance for the effects of binaries. 
Although Steele \& Jameson (1995) find that a photometric model of the I-band luminosity and I-K colour of artificial binary stars allowed the identification of a binary star sequence lying 0.75 mag above the single star sequence, the effect of an unresolved companion on the optical spectra would be small, shifting the spectral type by half a subclass at most. This is below the accuracy of the spectral types obtained from the spectra and therefore unmeasurable in this case. In this study, all the objects observed appear single in the images, therefore it is not possible to predict quantitatively the effects of unresolved binarity.

\subsection{Mass Segregation}
In such an extremely young cluster which is only a few crossing times old, we might expect that the members will not have had time to move far from their original positions of formation so dynamical mass segregation is not yet expected to be significant (e.g. de Grijs et al,  2002), except  for the most massive stars, which have become drawn to the cluster centre. There is some evidence that some 
mass segregation has already occurred for massive stars (Hillenbrand, 1997) but this will have a minimal effect on the low mass stars and brown dwarfs considered here.

\section{Summary and Conclusions}
\label{summary}
We have obtained optical spectra of a large number of low mass stars and brown dwarfs in the Trapezium Cluster using multi-object slit masks at the AAT and Gemini-North.  45 of the objects were assigned spectral types using molecular indices which measure the strength of highly temperature sensitive molecular features (see Paper I). These indices give spectral types insensitive to the presence of nebular emission lines or imperfect dereddening.  The spectra were also compared with the spectra of young Chamaeleon I objects to ensure correct spectral typing independent of luminosity class considerations. Spectral types were found to be in the range M3.25-M8.75.  An additional 11 objects showed basic M type characteristics but were unable to be typed accurately due to either low S/N or veiling. Spectral types obtained at both telescopes and on different runs were found to be in good agreement.

Cluster membership probability was investigated for all 45 typed objects by an analysis of the gravity-sensitive Na absorption feature near 8200\,\AA~. The strength of this feature was found to be much lower than that seen in older, higher surface gravity field dwarfs for all the sources except one, confirming their PMS nature and hence cluster membership. This form of evidence is extremely useful, being available and unequivocal in only moderate resolution spectra. Additional evidence was seen in the spectra of 6 objects, which showed strong CaII emission lines, indicative of accretion from a disk, also an indication of youth and hence cluster membership.  The depth of the Na absorption band appears to be a sensitive 
tracer of age in young clusters, and higher precision measurements may prove useful in investigating the age spread in some clusters. 

Spectral types were converted to effective temperatures using a temperature scale intermediate between those of dwarfs and giants, which is appropriate for young PMS objects (Luhman et al 2003b). The data were plotted on the HRD with dereddened H band magnitudes and compared with the evolutionary tracks and isochrones of Baraffe et al. (1998) and D'Antona and Mazzitelli (1997). For the majority of the sources, an average cluster age of 1 Myr was found, in agreement with several other studies.    However, the majority of the lowest mass members appear younger, often residing well above the 1 Myr isochrone.  While this could be a real feature of the cluster, it is more likely to be a result of selection effects in the sample observed.  We find a relative lack of an apparently older population at or near 10 Myr, as found by Slesnick et al. (2004). Instead we confirm the finding of Slesnick et al (2004) that at least some of  the apparently old sources are proplyds or other anomalously blue sources,  and suggest that they are younger than their positions in the HRD indicate.  

The results show that an average age of the Trapezium Cluster of 1 Myr is appropriate when estimating masses from photometric studies. They also show that masses estimated from photometry and spectroscopy are in generally good agreement, and MFs constructed from LFs are reliable. However, spectroscopy is still required to confirm cluster membership and to provide the most accurate estimates of mass, crucial for obtaining the true MF.

\section{Acknowledgments}

This work is based on observations obtained at the Anglo-Australian Observatory and  the Gemini Observatory, which is operated by the Association of Universities for Research in Astronomy, under a cooperative agreement with the NSF on behalf of the Gemini partnership: the National Science Foundation (United States), the Particle Physics and Astronomy Research Council (United Kingdom), the National Research Council (Canada), CONICYT (Chile), the Australian Research Council (Australia), CNPq (Brazil) and CONICET (Argentina).

We thank the Gemini and AAT Panels for the Allocation of Telescope Time for supporting this project, and the Gemini Observatory staff for carrying out Gemini observations for us. Many thanks also to Joss Bland-Hawthorn for invaluable advice and assistance before and during the AAT Taurus observations. 
We also thank Isabelle Baraffe for providing the theoretical isochrones. 

FCR acknowledges support by a PPARC doctoral studentship  and a Post-Doctoral Scholarship at Penn State University.

\section*{Appendix}

\appendix

\renewcommand{\thefigure}{A-\arabic{figure}}

\renewcommand{\thetable}{A-\arabic{table}}

\subsection*{A.1 Spectra}

\begin{figure*}
   \centering
   \resizebox{\hsize}{!}{\includegraphics[clip=true]{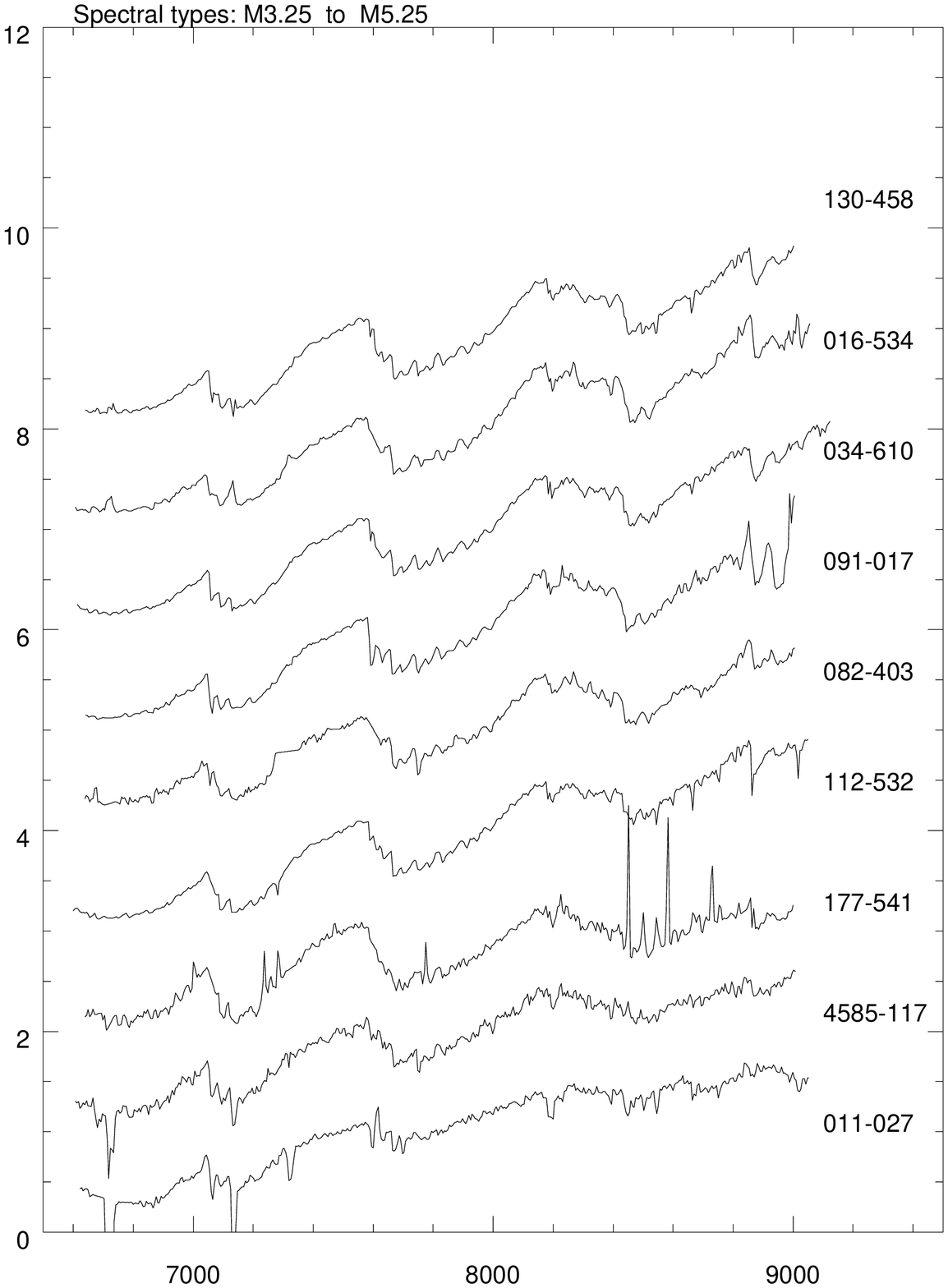} 
   \includegraphics[clip=true]{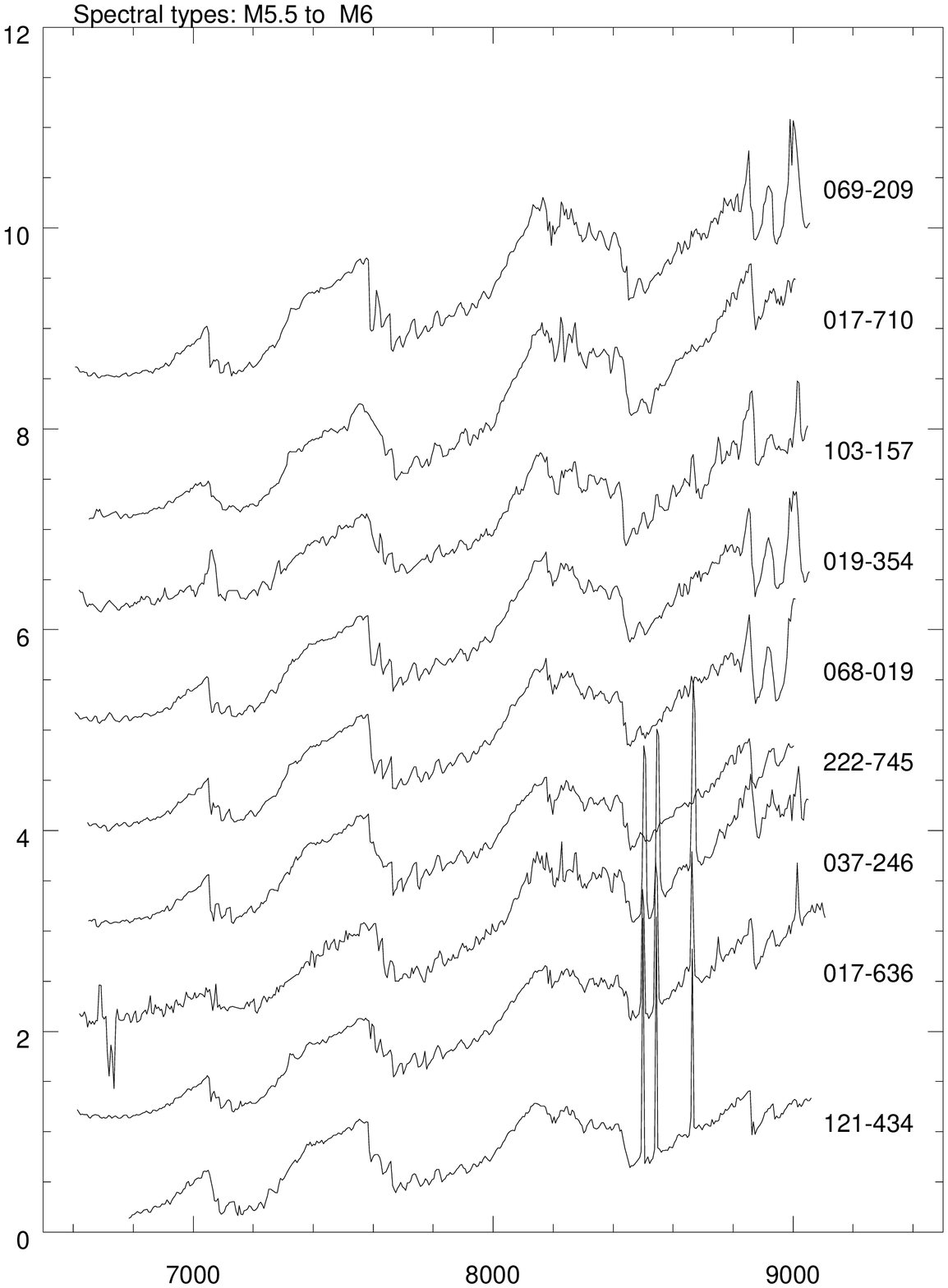}
   }
   
     \resizebox{\hsize}{!}{\includegraphics[clip=true]{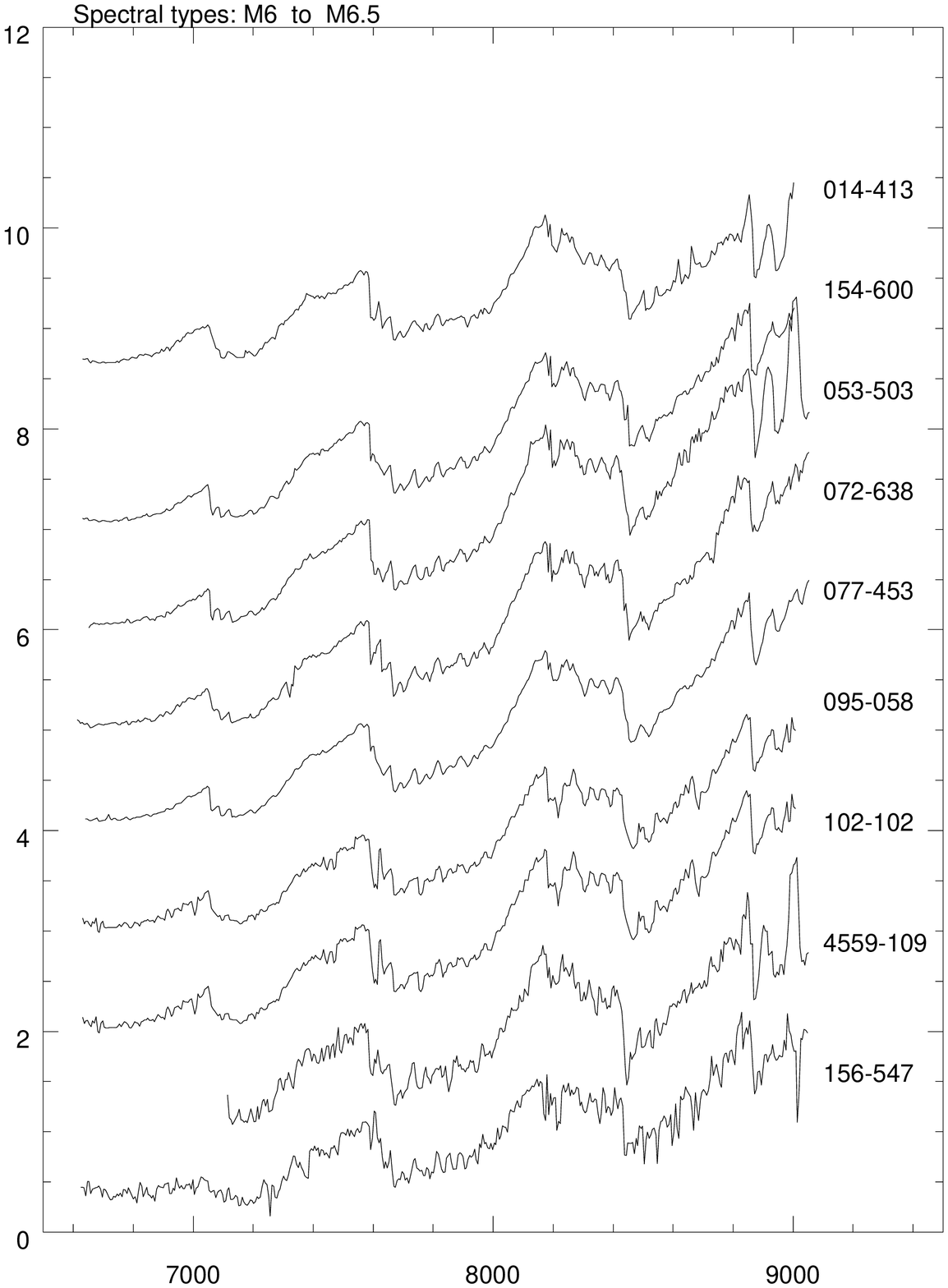} 
   \includegraphics[clip=true]{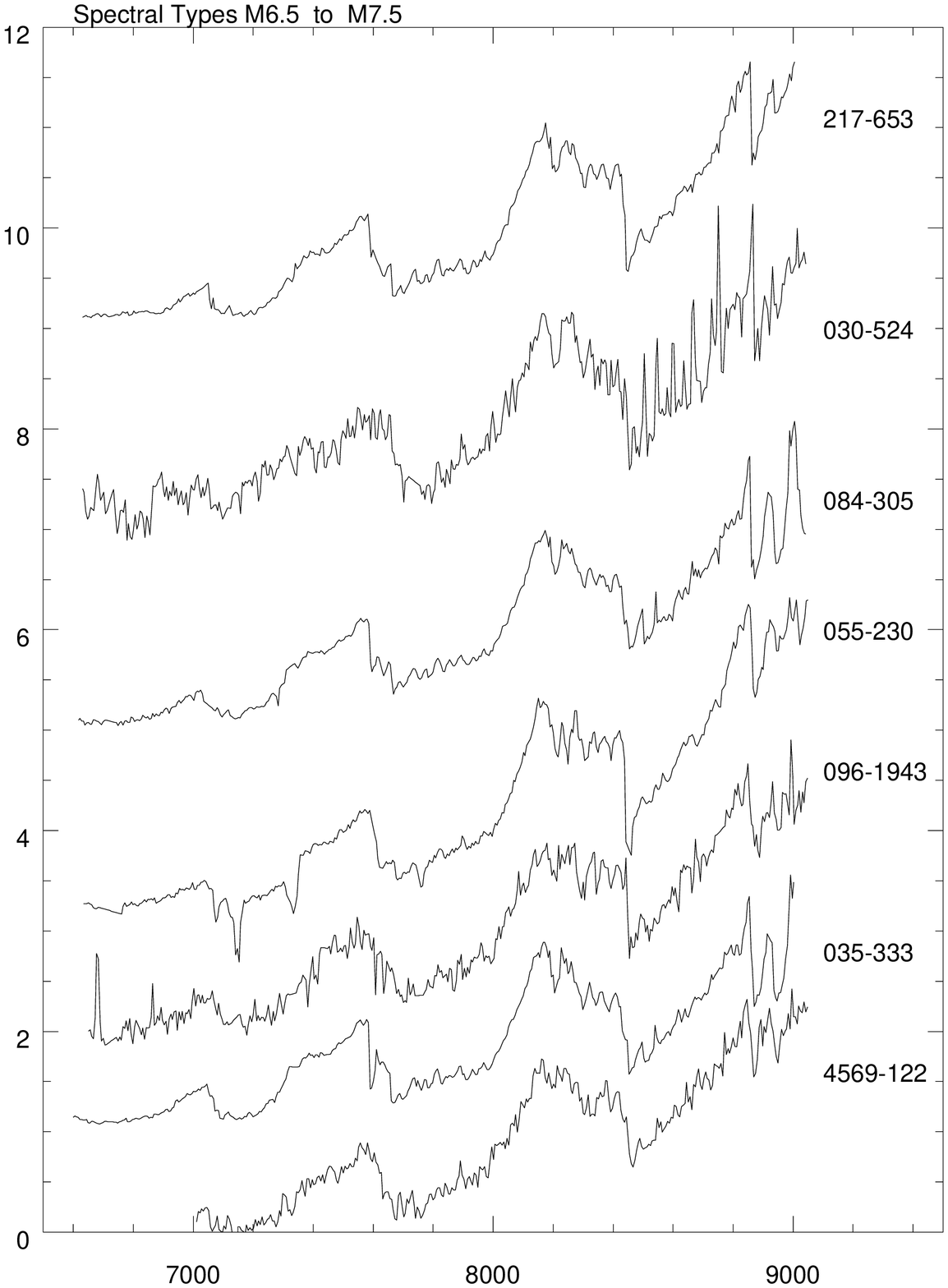}
 }
   
   \vspace*{-10mm}
     \caption{The 0.66 - 0.9~$\umu$m spectra of low mass stars and brown dwarfs spanning the range of spectral types in the sample.  The spectra are normalised at 0.75~$\umu$m  and displaced  vertically.  }
        \label{specplot}
    \end{figure*}

The 45 objects that have been classified are displayed in sequence of increasing spectral type in figure~\ref{specplot}.  The wavelength range displayed is  6600 to 9000\,\AA; the spectra  are contaminated by strong residual H$\alpha$ and [N II] emission below 6600\,\AA, while fringing and the falling CCD response reduces reliability above 9000\,\AA.  The spectra have been dereddened and smoothed to a spectral resolution of $\sim 10$\,\AA, and normalised at 7500\,\AA.   Errors in the dereddening applied will affect the slope of the spectra, but will have a smaller effect on the depth of the molecular absorption bands.   Residual nebular emission lines can be seen in emission or absorption, dependent upon the local emission gradient, in some objects, while circumstellar CaII triplet emission is present in 6 objects. 

Spectra of some of the objects that could not be classified reliably are shown in figure~\ref{specplot2}.  The three objects in the right panel show clear M-type spectral features, but with very strong residual nebular emission.    The residual nebular emission is positive in 046-245 and negative in 196-700.

The objects in the left panel are dominated by relatively featureless continuum emission, attributed to circumstellar material, along with residual nebular emission lines  and circumstellar CaII emission.  These objects suffer extreme amounts of veiling.  The spectra of 047-436 obtained at the AAT and Gemini are remarkable  in showing substantial spectral variations.  The spectral slope in the dereddened Gemini spectrum has a featureless continuum and rises to blue wavelengths, while the continuum  of the AAT spectrum is redder and displays evidence of M-type features. The CaII lines go from  emission in the Gemini spectrum to absorption in the AAT spectrum.  Some of these differences may be attributable to the different seeing conditions at the different observatories, but it is likely that there are real spectral variations too.  This type of spectral variability (correlated with photometric variations) has been observed in a more massive YSO (RR Tau) by Rodgers et al (2002).  Note that in these objects, the correct amount of reddening to be applied is very uncertain.   The amount of dereddening applied to 047-436, A$_{\rm V}$= 6.2 mag,  was derived from the photometry of LR00 and in view of the spectral variations and the circumstellar emission, must be regarded as highly uncertain.  There is only a single spectrum of 117-609, and here too the reddening estimate may not be reliable.

\begin{figure*}
   \centering
   \resizebox{\hsize}{!}{\includegraphics[clip=true]{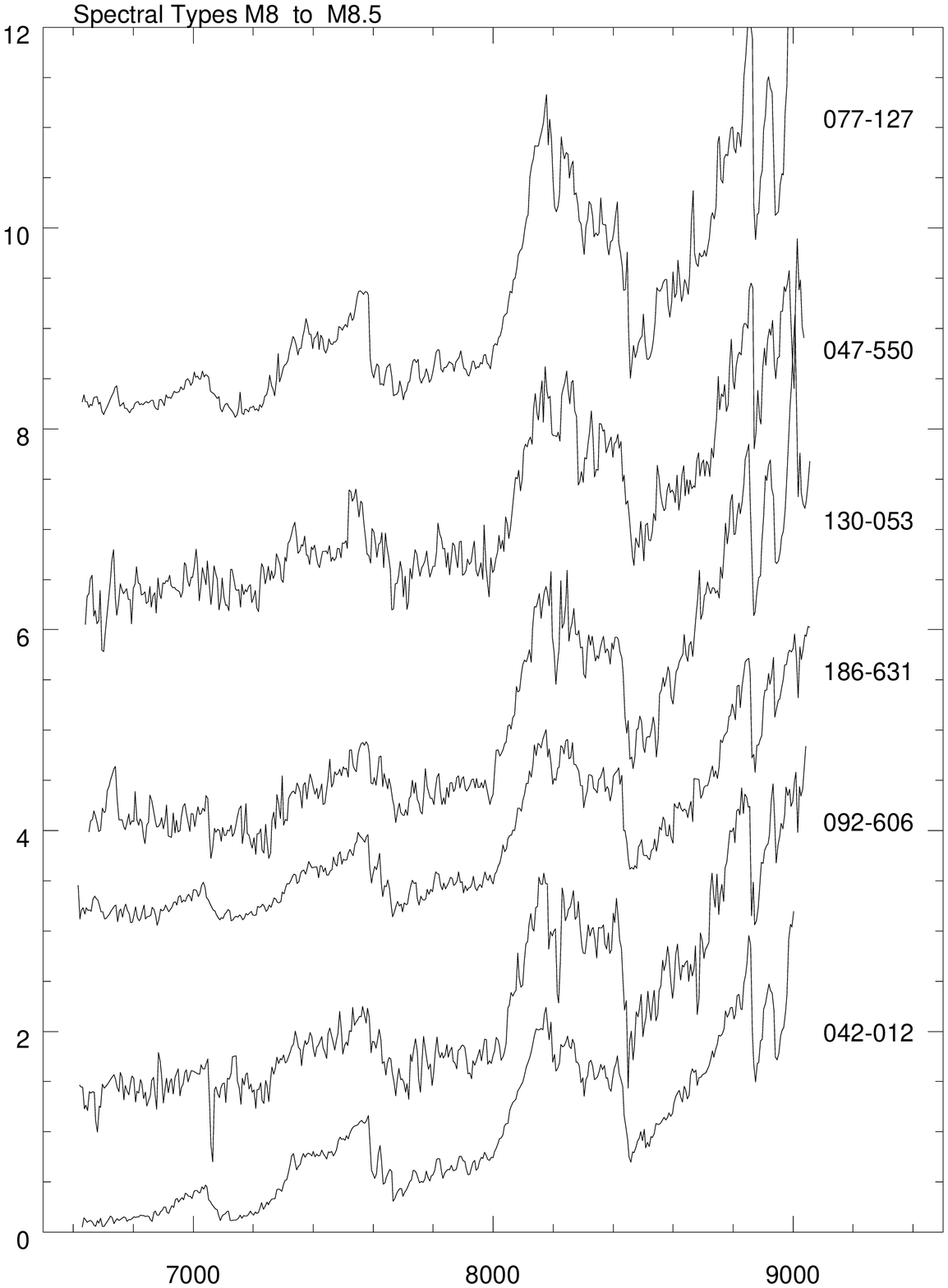} 
   \includegraphics[clip=true]{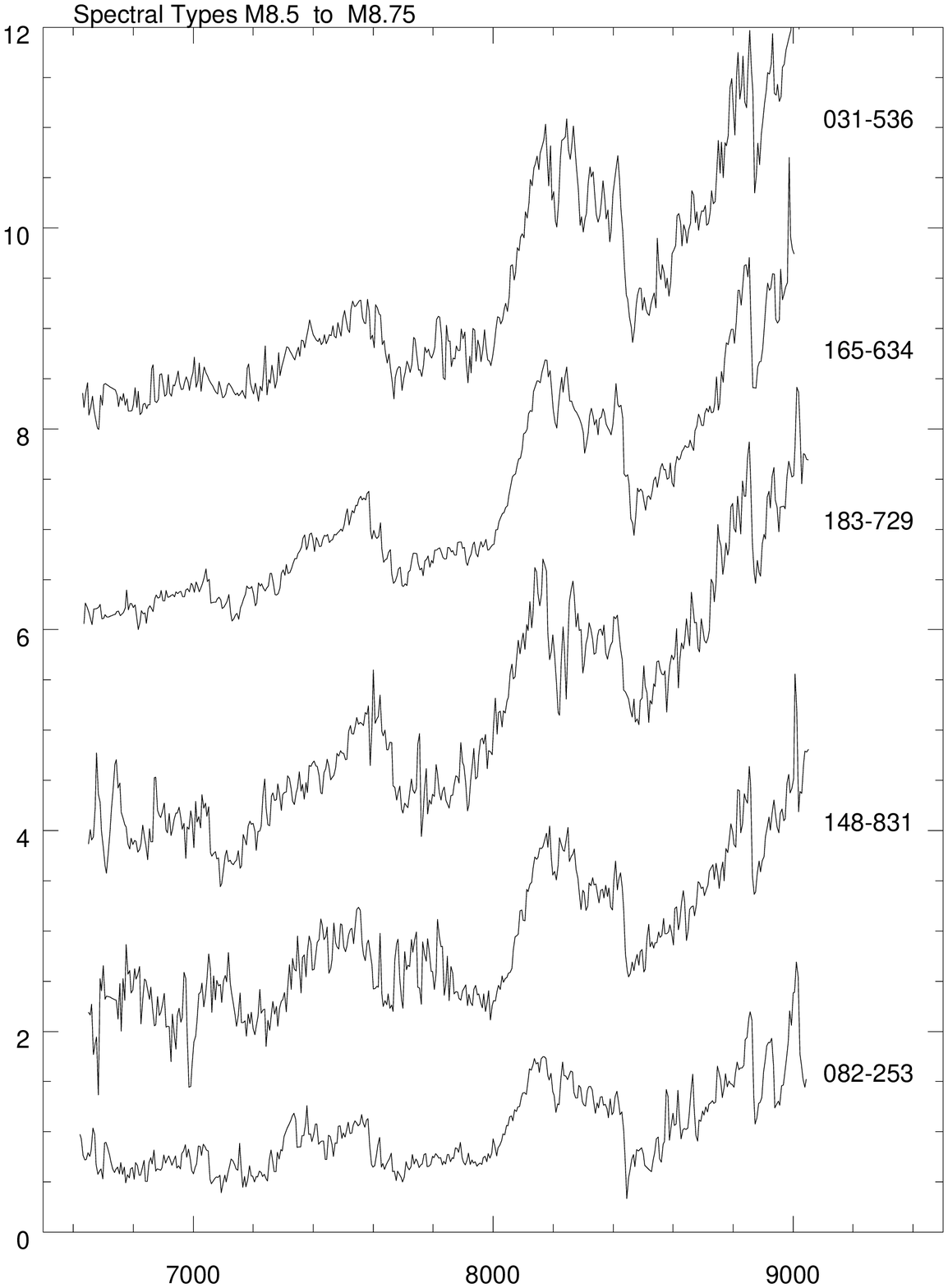}
 }
{Figure A-1 continued}
  \end{figure*}

\subsection*{A.2. Spectral Indices and Assigned Spectral Types}
The final spectral types and associated uncertainties are listed in Table A.1 for the objects that can be classified reliably.  Also shown are the spectral types given by the individual spectral indices recommended in paper I:  the VO 7445 index defined by Kirkpatrick et al (1991), the VO 2 index of  Lepine et al (2003), and the c81 index of Stauffer et al (1999).  The final classifications were decided manually, taking the indices into account.  The final column of the table lists the values of the Na index which provides surface gravity discrimination.

\begin{figure*}
      \centering
   \resizebox{\hsize}{!}{\includegraphics[clip=true]{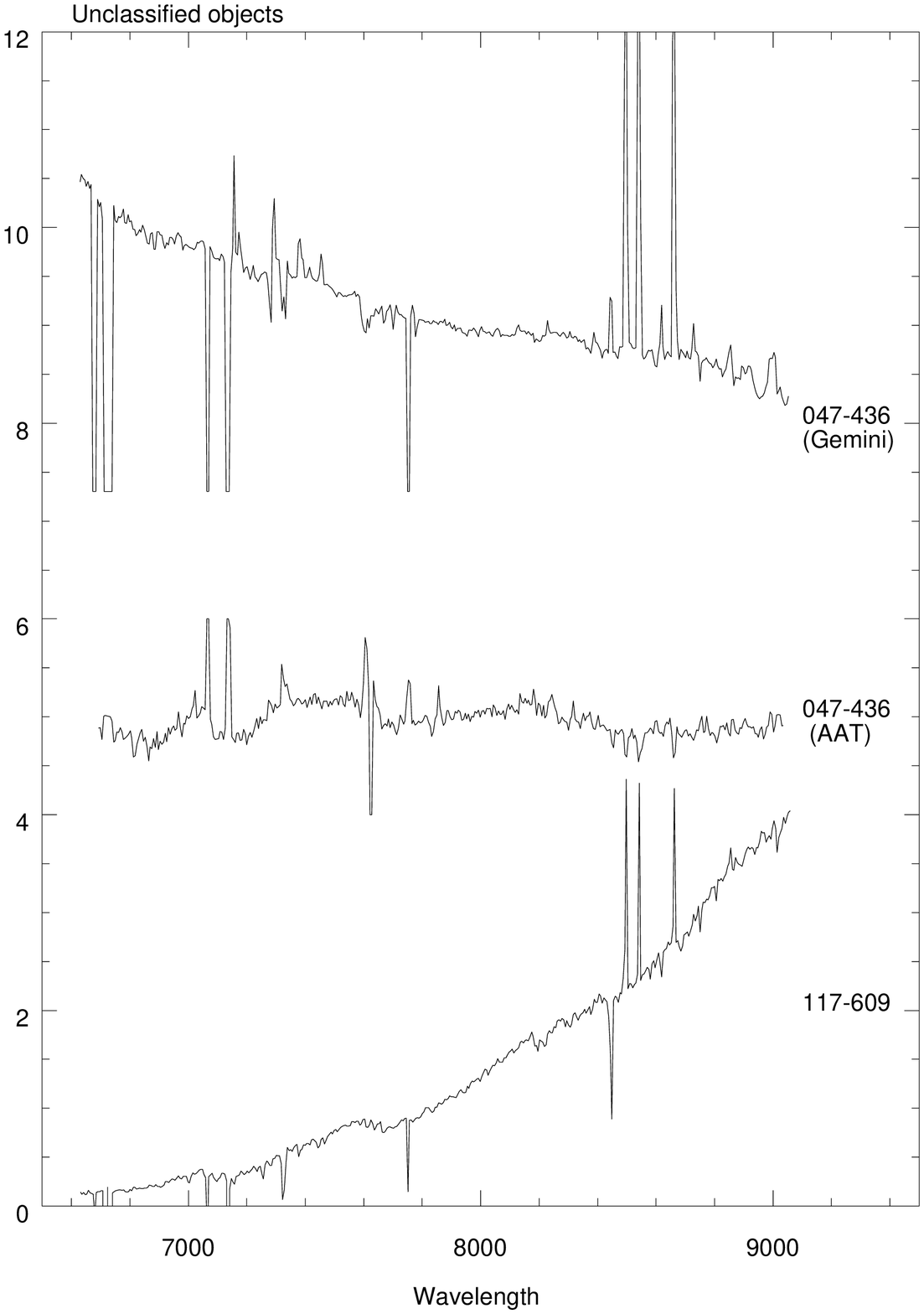} 
  \includegraphics[clip=true]{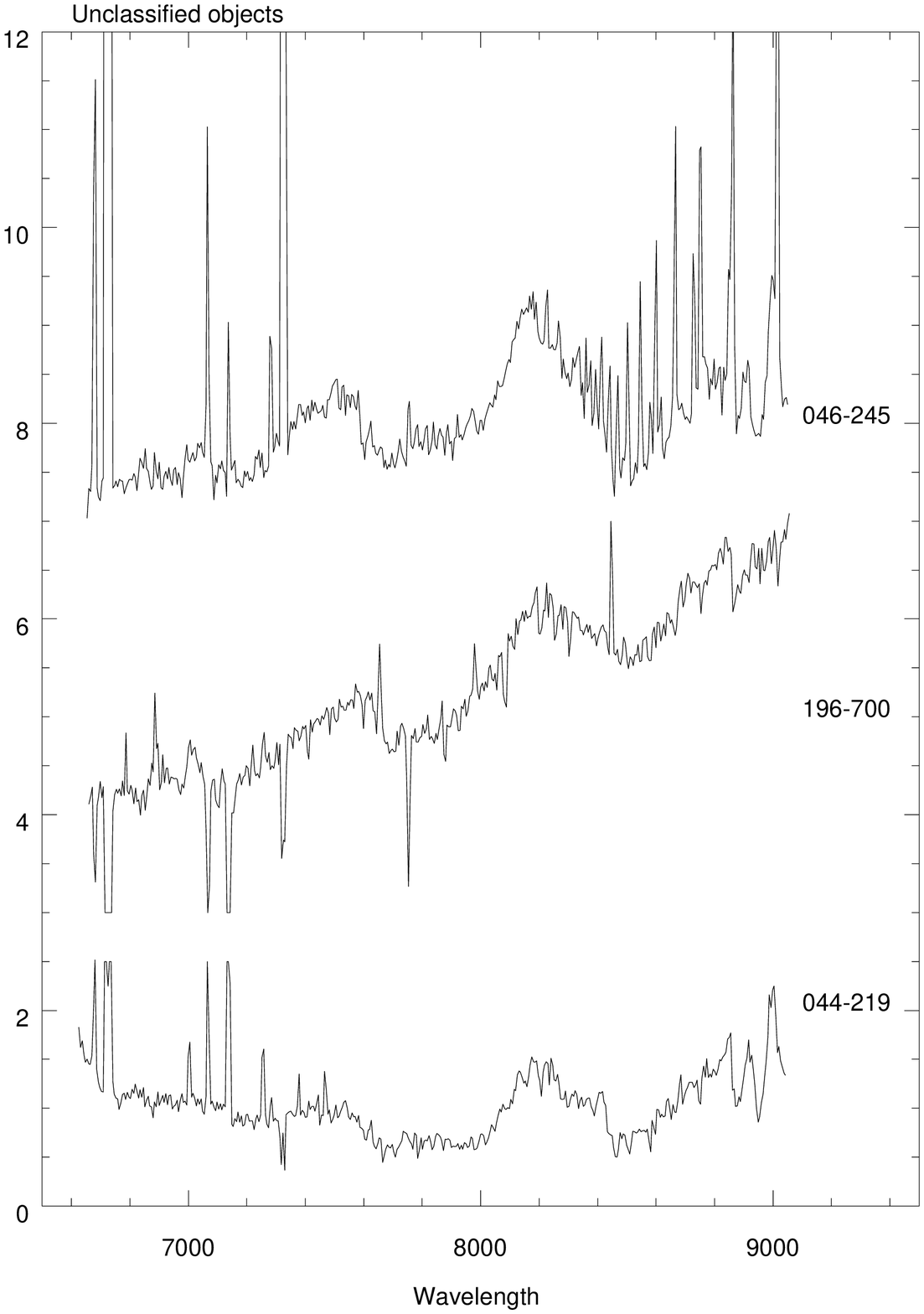}
   }

     \caption{Spectra of objects that could not be classified reliably.  Left: 2 objects with substantial veiling and CaII emission; note that the top two spectra are of the same object taken at different epochs. Right: 3 objects with M-type features but very strong residual nebular emission }
        \label{specplot2}
    \end{figure*}

\begin{center}
\begin{table*}
\caption{\textbf{Spectral Types from the individual and combined spectral indices }  }
\label{types}

\begin{tabular*}{390pt}{lcccccl}

\hline

Name &  Spectral Type & Type Error& Type (VO 7445) & Type (c81) & Type (VO2) & Na Index\\

\hline

011-027   & 3.25	& 1.0 & -- & 3.2 & 3.8 & 1.114\\
4584-117 &  3.50	& 0.5 & -- & 3.9 & 4.0 & 1.003\\
177-541 & 4.25	& 1.5 & -- & 4.4 & 4.3 & 0.961\\ 
112-532 & 4.75 	& 0.5 & -- & 4.6 & 5.2 & 1.008\\ 
082-403 & 4.75 	& 0.5 & -- & 4.8 & 4.8 & 0.980\\
091-017 & 5.00		& 0.5 & -- & 4.8 & 5.0 & 0.989\\
034-610 & 5.25	 & 0.5 & 5.5 & 5.0 & 5.1 & 0.987\\
016-534 & 5.25 & 0.5 & 5.3 & 5.2 & 5.4 & 0.998\\
130-458 & 5.25 & 0.5 & 5.4 & 5.3 & 5.3 & 0.99\\
121-434 & 5.50	& 1.0 & 5.3 & 4.6 & 5.2 & 0.975\\
017-636 & 5.50 & 1.0 & 5.6 & 3.6 & 4.5 & 0.973\\
037-246 &5.50 & 1.0 & $<$5.0~~ & 4.5 & 5.9 & 0.943\\ 
222-745 & 5.75	& 0.5 & 6.1 & 5.6 & 5.6 & 1.004\\
068-019 & 5.75	& 0.5 & 5.5 & 5.7 & 5.7 & 0.954\\
019-354 & 5.75 & 0.5 & 5.8 & 6.1 & 6.1 & 0.954\\
103-157 & 5.75 & 0.5 & 5.8 & 5.4 & 5.3 & 0.976\\
017-710 & 6.00 & 0.5 & 5.5 & 5.9 & 6.3 & 0.920\\
069-209 & 6.00 & 0.5 & 5.6 & 6.2 & 6.5 & 1.01\\
156-547 & 6.00	& 1.0 & 6.0 & 6.9 & 5.8 & 1.025\\
4559-109 &	 6.25	& 0.5 & 6.5 & 7.0 & 7.0 & 0.92\\
102-102 &	 6.25	& 0.5 & 6.6 & 5.9 & 6.4 & 1.01\\
095-058 &	 6.25	& 0.5 & 5.5 & 6.4 & 7.8 & 0.93\\
077-453 &	 6.50 & 0.5 & 5.9 & 7.3 & 7.0 & 0.919\\
072-638 &	 6.50 & 0.5 & 6.7 & 6.3 & 6.8 & 0.934\\
053-503 &	6.50 & 1.0 & 6.7 & 7.3 & 8.3 & 0.97\\
154-600 &	 6.50 & 0.5 & 6.2 & 6.6 & 6.7 & 0.931\\
014-413 &	 6.50 & 1.0 & 7.6 & 6.2 & 6.6 & 0.91 \\
4569-122 &	 6.75	& 1.0 & 5.0 & 7.5 & 7.8 & 0.865\\
035-333 &	 6.75 & 0.5 & 6.9 & 7.6 & 7.2 & 0.893\\
096-1943 &	 6.75	& 0.5 & -- & 6.7 & 6.8 & 0.881\\
055-230 &	 6.75 & 1.5 & 6.5 & 8.4 & 8.4 & 0.86\\
084-305 &         7.00	& 0.5 & 7.1 & 8.1 & 7.9 & 0.871\\
030-524 &	7.50 &1.5 & -- & 7.2 & 7.3 & 0.878\\
217-653 &	 7.75	& 0.5 & 7.2 & 8.0 & 7.7 & 0.883\\
042-012 &	7.75	& 1.0 & 7.5 & 8.3 & 7.9 & 0.862\\
092-606 &	 8.00	& 2.0 & 7.9 & -- & $>$M8.5 & 0.916\\
186-631 &	 8.00 & 1.0& 7.2 & 8.0 & 8.0 & 0.901\\
130-053 &	 8.50	& 2.0 & & -- & & 0.72\\ 
047-550 &	 8.50	& 1.5 & -- & 9.0 &$>$M8.5 & 0.829\\
077-127 &	 8.50	& 1.0 & 7.9 & 8.4 & 7.4 & 0.808\\ 
082-253 &	 8.50 &1.5 & 9.0 & 7.7 & 6.9 & 0.96\\
148-831 &	 8.50	& 2.0 & & & & 0.739\\
183-729 &	 8.75	& 1.5 & -- & 8.3 & 9.1 & 1.015 \\
165-634 &	 8.75 & 1.0 & $>$M8.5 & -- & $>$M8.5 & 0.797\\ 
031-536 &	 8.75 &1.0 & $>$M8.5 & $>$M8.5 & 8.4 & 0.84\\

\hline
\end{tabular*}
\end{table*}

\end{center}

\label{lastpage}

\end{document}